\documentclass{ar2e}
\usepackage{ulem}  
\usepackage{ARAstroBib}
\usepackage{color}
\pdfoutput=1

\usepackage{graphics}
\usepackage{epsfig}
\usepackage{amssymb}

\begin{document}



\jname{Annual Reviews of Astronomy and Astrophysics.}
\jyear{2011}
\jvol{1}
\ARinfo{1056-8700/97/0610-00}

\title{Solar Neutrinos: Status and Prospects}
\author{W. C. Haxton
\affiliation{ Department of Physics, University of California, Berkeley,  and Nuclear Science
Division, Lawrence Berkeley National Laboratory, Berkeley CA  94720-7300}
R. G. Hamish Robertson
\affiliation{Department of Physics, University of Washington,  and Center for Experimental
Nuclear Physics and Astrophysics, Seattle, WA 98195}
Aldo M. Serenelli
\affiliation{Instituto de Ciencias del Espacio (CSIC-IEEC), Facultad de Ciencias, 
Campus UAB, 08193 Bellaterra, Spain and
Max Planck Institute for Astrophysics, Karl Schwarzschild Str. 1, Garching, D-85471, Germany}}


\begin{abstract}
We describe the current status of solar neutrino measurements and of  the theory -- both neutrino
physics and solar astrophysics -- employed in interpreting measurements.  Important recent developments
include Super-Kamiokande's  determination of the $\nu-e$ elastic scattering rate
for $^8$B neutrinos to 3\%; the latest SNO global analysis in which the inclusion of low-energy
data from SNO I and II significantly narrowed the range of allowed values for the neutrino
mixing angle $\theta_{12}$; Borexino results for both the $^7$Be and pep neutrino fluxes, the first
direct measurements constraining the rate of ppI and ppII burning in the Sun;  global reanalyses
of solar neutrino data that take into account new reactor results on $\theta_{13}$; a new decadal 
evaluation of the nuclear physics of the pp chain and CNO cycle defining best values and uncertainties
in the nuclear microphysics input to solar models; recognition of an emerging discrepancy 
between two tests of solar metallicity, 
helioseismological mappings of the sound speed in the solar interior, and analyses of the metal
photoabsorption lines based on our best current description of the Sun's photosphere; a new round
of standard solar model calculations optimized to agree either with helioseismology or with
the new photospheric analysis;
and, motivated by the solar abundance problem, the development of 
nonstandard, accreting solar models, in order to investigate possible consequences of the metal
segregation that occurred in the proto-solar disk.  We review this progress and 
describe how new experiments such as SNO+ could help us further 
exploit neutrinos as a unique probe of stellar interiors.
\end{abstract}

\maketitle

\section{INTRODUCTION}
\label{sec:one}
In 1958  \cite{HJ58,HJ59} found that the cross section for ${}^3\mathrm{He}+{}^4\mathrm{He}\rightarrow {}^7\mathrm{Be}+\gamma$
was about 1000 times larger than anticipated, so that in addition to the simplest ${}^3\mathrm{He}+{}^3\mathrm{He}
\rightarrow {}^4\mathrm{He}+2\mathrm{p}$ ppI termination of the pp chain (see Fig. \ref{fig:cycles}), there might be significant branches to the ppII and ppIII cycles, 
and thus significant fluxes
of ${}^7$Be and ${}^8$B solar neutrinos.  Despite the uncertainties that existed in 1958 -- the solar core temperature was poorly constrained by theory,
and other nuclear physics important to the pp chain had not been resolved -- both \cite{Cameron}
and \cite{WAF58} pointed out that it might therefore be possible to detect solar neutrinos using a radiochemical
method Ray Davis had developed at Brookhaven \citep{Davis55}.  While the endpoint of the main source of neutrinos from the ppI cycle, 
p+p$\rightarrow$d+e$^+$+$\nu_e$, is below the 811 keV threshold for
$\nu_e$+${}^{37}$Cl$\rightarrow {}^{37}$Ar + e$^-$, most ${}^7$Be and $^8$B neutrinos are sufficiently energetic to
drive this reaction.  In 1962 Fowler organized a team of young Caltech researchers -- John Bahcall, Icko Iben, and
Dick Sears -- to begin the development of a solar model to more accurately predict the central temperature of the Sun and to estimate the rates
of neutrino-producing reactions \citep{BFIS}.   The history of these early developments is summarized in several sources \citep{Account,Lande10,Haxton10}. By early 1964, following significant advances in the solar model and in the understanding
of the nuclear physics of the pp chain and the ${}^{37}$Cl($\nu_e$,e$^-$)${}^{37}$Ar reaction, \cite{DavisPRL64} and \cite{JNBPRL64} concluded that a measurement of solar neutrinos would
be possible, were Davis to mount a detector 100 times larger than that he built at Brookhaven, in a site sufficiently deep to reduce backgrounds
from high-energy cosmic ray muons to an acceptable level.  In April 1968 \cite{DHH68} announced an upper bound on the 
solar neutrino capture rate for ${}^{37}$Cl of 3 SNU (1 SNU = 10$^{-36}$ captures/target atom/s), based on the initial running of a 100,000-gallon
C$_2$Cl$_4$ detector that the collaborators had constructed on the 4850-ft level of the Homestake gold mine, in Lead, South Dakota.

This upper bound, nearly a factor of three below the rate predicted by the \cite{BBS68} standard solar model (SSM),
began a controversy that  took 30 years to resolve.  Twenty of these years passed without independent confirmation of the Davis result:
as the Cl rate was a fraction of a count per day in 0.6 kilotons of organic liquid, other technologies 
with comparable sensitivity were not easily developed.  Because the Davis experiment was sensitive to a flux of neutrinos that varies
sharply with the solar core temperature ($\phi(^8$B) $\sim T_C^{22}$ \citep{Book}), the result could be accommodated by a variety of possible
changes in the SSM having
the net effect of reducing $T_C$ by $\sim$ 5\%.   But as additional constraints on solar neutrino fluxes were established by the Kamiokande \citep{KamiokandeA}, SAGE \citep{SAGE}, and
GALLEX \citep{GALLEX} collaborations, a more detailed pattern of fluxes
emerged that was not easily reconciled with the SSM.
In contrast, with the discovery of the MSW mechanism \citep{MS1985,MS1986,Wolfa, Wolfb},
it became apparent that neutrino oscillations augmented by matter effects could account for the observations, 
even for a small mixing angle.  The conclusions of an Annual Reviews article from this period \citep{Haxton95}
captures the sense of excitement that with new experiments, a resolution of the solar neutrino problem was near.

 In 1998 $\nu_\mu \rightarrow \nu_\tau$ vacuum neutrino oscillations were discovered through the Super-Kamiokande collaboration's 
study of the zenith-angle dependence of atmospheric neutrino fluxes \citep{SK98}.  While this result did not directly constrain the $\nu_e$s produced by
the Sun, the discovery was a game-changer, confirming a phenomenon originally suggested by \cite{Pontecorvo67}
as a possible explanation for the solar neutrino problem.  Finally, the Sudbury Neutrino Observatory (SNO)
collaboration \citep{SNO1,SNO2} measured both the $\nu_e$ and heavy-flavor components of the solar neutrino flux arriving at Earth,  utilizing three
different detection channels with varying sensitivities to charge and neutral currents.  The SNO collaboration
measured the electron and heavy flavors components of the $^8$B solar neutrino flux,
found that the total flux of neutrinos (summed over flavors) is in good agreement with the SSM prediction, and determined flavor mixing parameters
that attributed the differential suppression of the pp, $^7$Be, and $^8$B fluxes deduced from previous experiments to the energy-dependent
effects of matter on oscillations.

This review summarizes the basic physics of solar neutrinos, the work that led to the discoveries
noted above, and the impact of recent and ongoing solar neutrino experiments on
astrophysics and weak interactions, including
\begin{itemize}
\item Completion of phase III of the Super-Kamiokande experiment \citep{SKIII}
and preliminary results from Super-Kamiokande IV's low-threshold running \citep{Smy};
\item SNO's combined analysis of all three SNO phases \citep{SNOCombined} and the collaboration's
low-energy analysis of the data from SNO I and II \citep{SNOLET};
\item  Borexino's achievement of a 5\% measurement of the $^7$Be flux, an initial result for the pep flux, and
a limit on the CN neutrino contribution \citep{Borexino,Borexinopep}; and
\item Daya Bay, Reno, and Double Chooz measurements of $\theta_{13}$, impacting global 
analyses of
solar neutrino data \citep{DayaBay,Reno,DC}.
\end{itemize}
In addition, a comprehensive survey
of the nuclear physics of the pp chain and CNO cycle has been completed, yielding a new set of best values and uncertainties for 
the nuclear rates \citep{SFII}.  The sound speed throughout most of the Sun's interior has been
extracted from helioseismology to an accuracy $\sim$ 0.1\%, providing a stringent check on SSM
predictions.
More sophisticated 3D models of the solar atmosphere have been developed, significantly improving the agreement between predicted and
observed absorption line-shapes and the consistency of abundance determinations from different atomic and molecular lines  \citep{AGSS09} -- but also yielding a photospheric metal abundance $\sim$ 30\% below 1D values, leading
to a conflict between SSMs employing the new abundances and solar parameters deduced from
helioseismology. The SSM has been updated
for the new nuclear reaction rates and alternative
metallicities, and nonstandard models have been developed to explore accretion as a possible
solution to the ``solar abundance problem" \citep{Serenelli09,Serenelli11}.  

For three decades solar neutrino physics was defined by an incomplete knowledge of the neutrino
fluxes and shortcomings in our understanding of neutrino flavor physics.
We are now starting a new period, where precise spectroscopy of the full spectrum
of solar neutrinos is possible, and where a clearer understanding of weak interactions has
been obtained from a combination of astrophysical, reactor, and accelerator experiments.
On one hand, this returns us to the roots of solar neutrino
physics: with weak interaction uncertainties removed, solar neutrinos can be used
to probe possible limitations in the SSM -- such as
uncertainties in the Sun's initial composition and the absence of {\it ab initio} treatments of mixing and
other three-dimensional physics, including rotation and magnetic fields.  On the other hand,
the neutrinos coming from the Sun remain important to fundamental physics:
the spectral shapes and fluxes of the various sources are  known rather precisely, and low-energy 
neutrinos react with targets rather simply, giving us
confidence that we can interpret measurements.
Thus this review also considers the continuing role solar neutrinos could play
in further constraining the Pontecorvo-Maki-Nakagawa-Sakata (PMNS) neutrino mass
matrix and in probing matter effects and other environmental neutrino phenomena.

\section{THE SSM AND ITS NUCLEAR AND NEUTRINO PHYSICS}
\label{sec:two}
\subsection{The Standard Solar Model}
Solar models trace the evolution of the Sun from its beginning -- when the collapse of the pre-solar
gas cloud was halted by the turn-on of thermonuclear reactions -- to today,
thereby predicting contemporary solar properties such as the composition, temperature, 
pressure, and sound-speed profiles and the neutrino fluxes.  SSMs are
based on four assumptions \citep{Book}:
\begin{itemize}
\item The Sun burns in hydrostatic equilibrium, maintaining a local balance between the
gravitation force and pressure gradient.  To implement this condition, an equation of state (EoS)
is needed.  As the hydrogen and helium in the Sun's core are nearly completely ionized, an ideal gas EoS
with corrections for incomplete ionization of metals, radiation pressure, and screening is thought to provide a
good approximation to the EoS \citep{Book}.  Helioseismic inversions of solar p-mode frequencies
have provided important tests of our understanding of the associated theory \citep{Elliott98}.
\item The mechanisms for energy transport are radiation and convection.  The inner portion
of the Sun -- 98\% by mass or about 71\% by radius -- is radiative.  In order to describe radiative
transport, the opacity must be known as a function of temperature, density, and composition.  
In addition to elementary processes such Thomson scattering off electrons and inverse bremmstrahlung off
fully ionized hydrogen and helium, more complex processes such as bound-free scattering off metals are important 
contributors to the opacity in the Sun's central regions.  Thus modern opacity tables like OPAL
\citep{Rogers02} are based on detailed atomic input.
Changes in opacity influence important helioseismic properties such as the sound speed and
the location of the convective zone boundary.
In the Sun's outer envelope, where the radiative gradient is larger, 
convection dominates the energy transport.  
SSM convection 
is modeled through mixing length theory, in which volume
elements are transported radially over a characteristic distance determined empirically in
the model, but typically on the order of the pressure scale height.
\item The Sun produces its energy by fusing protons into ${}^4$He,
\begin{equation}
2 e^-+ 4 \mathrm{p} \rightarrow {}^4 \mathrm{He}  + 2 \nu_e +\mathrm{26.73~MeV}.
\label{eq:4p}
\end{equation}
via the pp-chain ($\sim$ 99\%) and CN cycle ($\sim$ 1\%).  The nuclear cross sections are
taken from experiment or, if that is impractical, from theory: the associated
laboratory astrophysics is challenging because reaction rates are needed for
energies well below the Coulomb barrier.  Thus laboratory
measurements are generally made at higher energies, with theory guiding the
extrapolations to the solar Gamow window.
\item Boundary conditions include the modern Sun's mass, age, radius $R_\odot$, and luminosity $L_\odot$.  The 
pre-solar composition
is divided into hydrogen X$_\mathrm{ini}$, helium Y$_\mathrm{ini}$, and metals Z$_\mathrm{ini}$, 
with X$_\mathrm{ini}$+Y$_\mathrm{ini}$+Z$_\mathrm{ini}$=1.
Relative metal abundances can be determined from a combination
of photospheric (determined from analyses of absorption
lines) and meteoritic (for refractory elements) abundances, and are generally assumed to have
remained constant since the Sun formed.  The photospheric abundances
and the assumption of a homogeneous zero-age
Sun then constrain the Sun's initial core composition:  one can equate the Sun's pre-solar
core metallicity Z$_\mathrm{ini}$ to its present photospheric Z$_S$, corrected for 
the effects of diffusion over the
Sun's lifetime.   Finally Y$_\mathrm{ini}$ and the mixing length $\alpha_\mathrm{MLT}$ are determined
interatively by demanding that $L_\odot$ and $R_\odot$ are produced 
after 4.6 Gy of burning.
\end{itemize}
\noindent
The resulting model is dynamic.  The luminosity of the Sun increases by $\sim$ 40\%
over the solar lifetime: Helium synthesis alters the mean molecular weight and opacity in the core, requiring
a response in $T_C$.  The ratio of ppI/ppII/ppIII burning changes,
with the fraction of energy produced through the more temperature-sensitive ppII and
ppIII branches increasing.  The ${}^8$B neutrino flux for the ppIII cycle 
has an exceedingly sharp growth $\sim e^{t/\tau}$ where $\tau \sim$ 0.9 Gy.
Composition gradients are created as the pp chain burns to
equilibrium.  An interesting example is the solar core $^3$He abundance, which rises steeply with
increasing radius,
$X_3 \propto T^{-6}$, throughout the region where pp-chain equilibrium has been reached.
The $^3$He abundance gradient was proposed as a potential trigger for periodic mixing of the core in the
``solar spoon" \citep{DilkeGough}.   Metals are rearranged: in the first 10$^8$ years of main-sequence burning
most of the carbon in the Sun's central core is converted to nitrogen, building up the core
abundance of $^{14}$N.  Because 
they have a smaller charge-to-mass ratio than hydrogen, $^4$He and metals
slowly diffuse toward the core -- another source of composition gradients that 
affect contemporary observables like helioseismology.  

\begin{table}
\caption{SSM characteristics are compared to helioseismic
values, as
determined by \cite{Basu97,Basu04}. X, Y, and Z are the mass
fractions in H, He, and metals. The subscripts $S$, $C$, and ini denote current
photospheric, current core, and zero-age values.  $R_{CZ}$ is
the radius to the convective zone,
and $\langle \delta c/c \rangle$ is the average fractional discrepancy in 
the sound speed, relative to helioseismic values.}
\vspace{0.4cm}
\label{tab:SSM}
\begin{tabular}{lccc}
\hline \hline
Property & GS98& AGSS09& Solar \\
(Z/X)$_S$ & 0.0229 & 0.0178 & -- \\
Z$_S$ & 0.0170 & 0.0134 & -- \\
Y$_S$ & 0.2429 & 0.2319 & 0.2485$\pm$0.0035 \\
$R_{\rm CZ}$/$R_\odot$ & 0.7124 & 0.7231 & 0.713$\pm$0.001 \\
$\left< \delta c / c\right>$ & 0.0009 & 0.0037 & 0.0\\
Z$_C$ & 0.0200 & 0.0159 & -- \\ 
Y$_C$ & 0.6333 & 0.6222 & --\\ 
Z$_{\rm ini}$ & 0.0187 & 0.0149 & -- \\
Y$_{\rm ini}$ & 0.2724 & 0.2620 & -- \\
\hline \hline
\end{tabular}
\end{table}

Properties of two SSMs we will use in this review are listed in Table \ref{tab:SSM}. 
The models differ in the values assumed for the
photospheric metallicity Z$_S$, with the GS98-SFII SSM
being more metal rich than the AGSS09-SFII SSM.  The table gives the model photospheric
helium Y$_S$ and  metal Z$_S$ abundances, the radius
of the convective zone $R_\mathrm{CZ}$, the mean deviation of the sound speed $\left< \delta c / c\right>$
from the helioseimic profile, the core helium and heavy element abundances
Y$_C$ and Z$_C$, and the Sun's pre-solar abundances
Y$_\mathrm{ini}$ and Z$_\mathrm{ini}$.

\subsection{The pp Chain and CN cycle}
Approximately 80\% of the observed stars lie along a path in the Hertzsprung-Russell
diagram characterized by energy generation through proton burning.
The Sun provides a unique opportunity to test our understanding of such main-sequence
stars, as we can compare model predictions to solar properties that are 
precisely known.  This has inspired a great deal of laboratory
work to reduce uncertainties in atomic opacities and nuclear cross sections -- key SSM input parameters --
so that we can assess the reliability of the more fundamental solar physics and weak interactions
aspects of the model.

\begin{figure}
\begin{center}
\includegraphics[width=14cm]{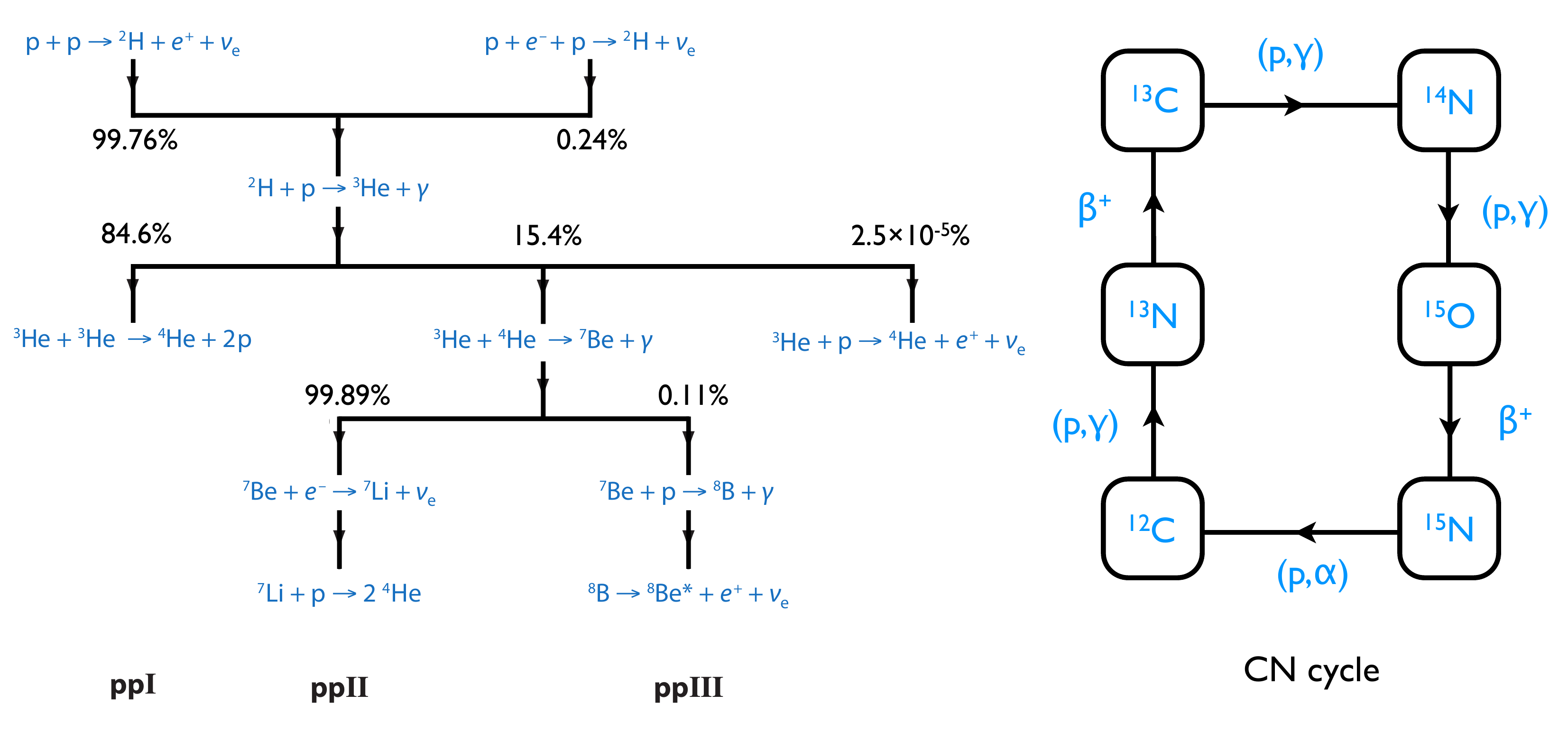}
\caption{(Color online) The left frame shows the three principal cycles comprising the pp chain (ppI, ppII, and ppIII), the associated neutrinos that ``tag" each of the three branches, and the theoretical
branching percentages defining the relative rates of competing reactions
(GS98-SFII SSM).
Also shown is the minor branch ${}^3$He+p $\rightarrow$
${}^4$He+e$^+$+$\nu_e$, which generates the most
energetic neutrinos.  The right frame shows the CN I cycle, which produces
the $^{13}$N and $^{15}$O neutrinos.} 
\label{fig:cycles}
\end{center}
\end{figure}


In lower-mass hydrogen-burning stars $^4$He is synthesized primarily through the pp-chain
nuclear reactions shown on the left side of Fig. \ref{fig:cycles}.  
The rates of the ppI, ppII, and ppIII 
cycles comprising the pp chain can be determined from the fluxes of
the pp/pep, $^7$Be, and $^8$B neutrinos produced by those cycles.
As the relative rates are very sensitive to $T_C$,
the neutrino fluxes are a sensitive thermometer for the solar core,
provided the associated nuclear physics is under control.

Rates depend on the quantity $\langle \sigma v \rangle_\mathrm{MB}$ where $v$ is the 
relative velocity between two colliding nuclei, $\sigma$ is the cross section, and $\langle ~\rangle_\mathrm{MB}$ denotes an
average over the Maxwell-Boltzmann relative velocity distribution in the solar
plasma.  The optimal energy for a solar reaction, called the Gamow peak, is determined from the
competition between a cross section that rises rapidly as the energy climbs the Coulomb barrier,
and a relative-velocity distribution that declines rapidly on the Maxwell-Boltzmann tail. 
Two ppI-cycle reactions, $d+p$  and
${}^3$He+${}^3$He, have been measured in their Gamow peaks.
Data were obtained down to 2.5 and 16 keV, respectively, at LUNA, Gran Sasso's
low-background facility for nuclear astrophysics \citep{Bonetti99}.
For other pp-chain reactions, direct measurements are not currently possible because of the severity of
the Coulomb suppression.
Instead, measurements must be made at higher energies, then extrapolated
to solar energies using nuclear models to predict the cross section
shape.

Such extrapolations are usually performed by using the S-factor, 
\begin{equation}
\sigma \left( E \right) =\frac{\mathrm{S} \left( E
\right)}{E}\;\;\mathrm{exp}\left[ {-2\pi \eta(E) } \right],
\label{eq:S}
\end{equation}
which removes from the cross section
the rapid energy dependence associated with the s-wave interaction
of point nuclei.  Here the Sommerfeld parameter 
$\eta(E)$ = $Z_1 Z_2$ $\alpha$/$v$, with 
$Z_1$ and $Z_2$ the ion charges, the relative velocity $v=\sqrt{2E/\mu}$,
$\mu$ is the reduced mass, and
$\alpha$ is the fine structure constant
($\hbar$ = c = 1).   The remaining nuclear physics (including effects of finite nuclear size,
higher partial waves, antisymmetrization, and any atomic screening
effects not otherwise explicitly treated) is then isolated in S(E), the function used in
extrapolations because of its gentler dependence on E.  In solar applications
S(E) is frequently approximated 
by its zero-energy value S($0$) and corrections determined
by its first and second derivatives, S$^{\prime}(0)$ and
S$^{\prime \prime}(0)$.  

\begin{figure}
\begin{center}
\includegraphics[width=13cm]{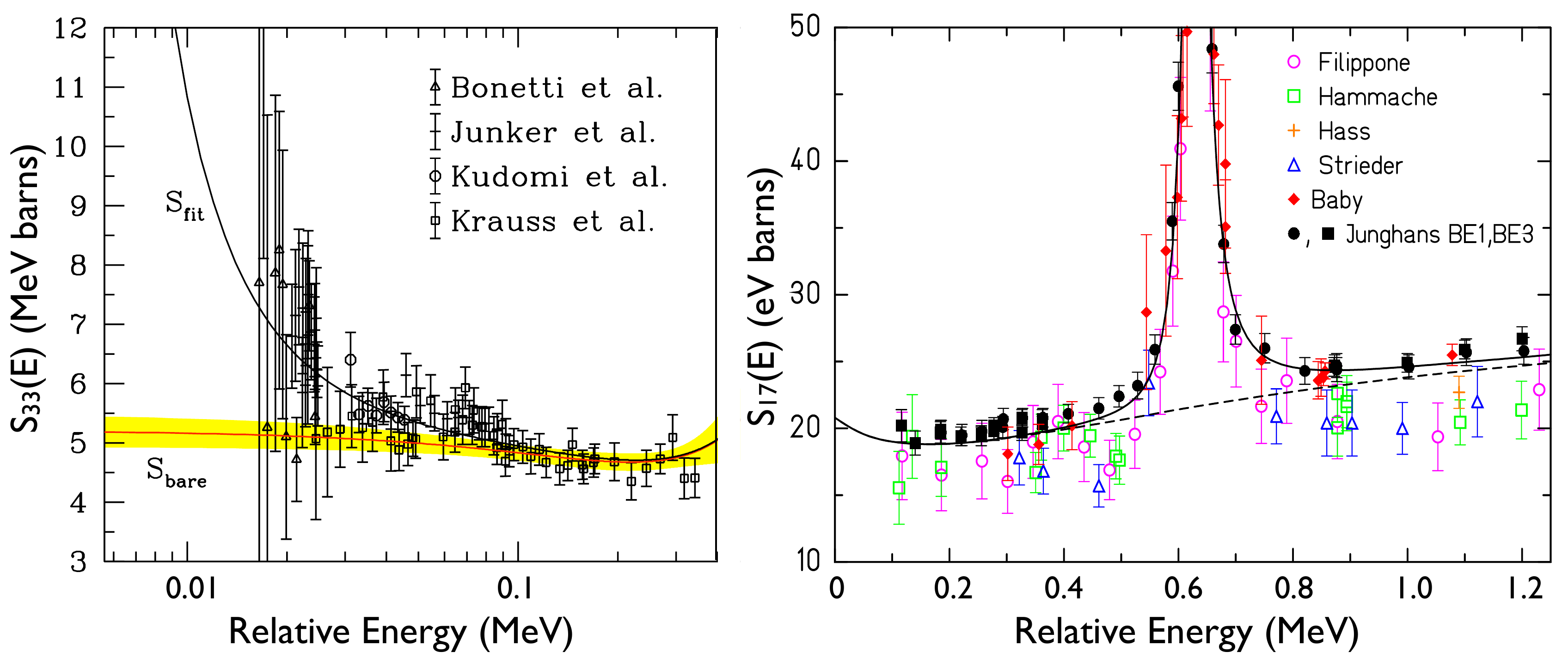}
\caption{(Color online) Left panel: The data, the best quadratic+screening result for S$_{33}(E)$,
and the deduced best quadratic fit (red line) and allowed range (yellow band) for S$^\mathrm{bare}_{33}$.  
Right panel: S$_{17}(E)$ vs. center-of-mass energy $E$, for $E \leq$ 1250 keV. 
Data points are shown with total errors, including systematic errors.  The dashed line
is based on a theoretical calculation, scaled to fit the data.  See Solar Fusion II for references
and details.}
\label{fig:Sfactors}
\end{center}
\end{figure}

The recent review by \cite{SFII} (Solar Fusion II) details the data and theory issues
affecting our understanding of solar cross sections.
Uncertain S-factors remain one of the key limitations in SSM neutrino flux predictions. 
Figure \ref{fig:Sfactors} gives the Solar Fusion II summaries for
${}^3$He+${}^3$He$\rightarrow^4$He+2p (left panel)
and ${}^7$Be+p$\rightarrow^8$B+$\gamma$ (right panel).  While measurements for the first reaction
cover the solar energies of
interest, the screening environment
of a terrestrial target (neutral atoms) differs from that at the center of the Sun
(ionized ${}^3$He). It is apparent from  Fig. \ref{fig:Sfactors} that the theoretical screening correction is significant.

The reaction $^7$Be(p,$\gamma)^8$B (right panel of Fig. \ref{fig:Sfactors}) feeds  the ppIII cycle that produces
the $^8$B neutrinos measured by SNO and Super-Kamiokande.   This reaction was considered the 
most uncertain in the pp chain when these detectors began operations, with only a
single data set considered sufficiently reliable and well 
documented for inclusion in an S-factor determination \citep{SFI}.
Four new, high-quality data sets were available for the Solar Fusion II evaluation, yielding
 S$_{17}$(0)=20.8  $\pm$  0.7  (expt)  $\pm$  1.4(theor).   The dominant error
 is now the theoretical extrapolation to the Gamow peak.

The CN I cycle, illustrated on the right in Fig. \ref{fig:cycles}, is typically the dominant mode for hydrogen burning
in massive main-sequence stars, where core temperatures exceed those of the Sun.  Unlike the pp chain, the CN cycle
depends on pre-existing metals to catalyze a series of (p,$\gamma$) and (p,$\alpha$) reactions,
leading in sum to Eq. (\ref{eq:4p}).  Thus the CN cycle is (in most settings) a secondary
process, dependent on the star's metallicity.  
In the Sun the CN cycle converts C to N as it seeks equilibrium.  Equilibrium has been reached
only in the most central regions of the core, where
$T \gtrsim 1.33 \times 10^7$K.  Outside this
region, the bottleneck reaction
$^{14}$N(p,$\gamma$) inhibits cycling.   Thus, throughout most of the cooler regions of the core, very little CN-cycle burning
takes place: pre-solar $^{12}$C has already been converted to
$^{14}$N, but little $^{14}$N is being consumed.  
Still further outward,  where $T \lesssim 10^7$K, the $^{12}$C lifetime is
comparable to the solar age.  This is the region in the contemporary Sun where pre-solar $^{12}$C
is being burned.   Thus CN-cycle neutrinos, produced in the $\beta^+$ decay
of $^{15}$O and $^{13}$N, come from two distinct regions.  Deep in the solar core equal 
numbers of $^{15}$O and $^{13}$N neutrinos are
produced in equilibrium burning, while in the distant cool outer regions of the core, 
only $^{13}$N neutrinos are produced.

\subsection{Solar Neutrino Fluxes}
The main neutrino-producing reactions of the pp chain and CN cycle are summarized in
Table \ref{tab:fluxes}.  Four of the five $\beta$ decay reactions produce neutrino spectra with allowed shapes and endpoints given by E$_\nu^\mathrm{max}$.   In the
fifth, the decay of $^8$B, the $^8$Be final state is a broad ($\sim$ 2 MeV) resonance.  As
the profile of this resonance is known, the associated small deviations from an allowed
spectrum can be calculated.  In addition to the main pp/pep, $^7$Be, and $^8$B neutrinos,
a fourth source from a weak side-branch of the pp chain, the hep or $^3$He+p neutrinos, 
is shown in Fig. \ref{fig:cycles}.  These neutrinos are the most energetic produced by the Sun
(E$_\nu^\mathrm{max}$ $\sim$ 18.77 MeV), and thus may be observable in SNO and
Super-Kamiokande they populate energy bins above
the $^8$B neutrino endpoint.
The two electron-capture (EC) reactions, p+e$^-$+p and $^7$Be+e$^-$, produce line sources of neutrinos
of energy E$_\nu^\mathrm{max}$, broadened by
thermal effects.  
There are two lines associated with $^7$Be EC, as $\sim$ 10\% of the captures populate the
first excited state in $^7$Li.

The ppI, ppII, and ppIII contributions to solar energy generation
can be determined from measurements of the pp/pep, $^7$Be, and $^8$B
neutrino fluxes.
As we will discuss in the next section, the $^7$Be and $^8$B fluxes are now quite well known,
while the first measurement of pep neutrinos was very recently announced.  
The ``solar values" of Table \ref{tab:fluxes} come from the Borexino
Collaboration (private communication) which updated an analysis by \cite{BahcallPG}, combining
$^8$B, $^7$Be, and pep flux measurements (as available in March, 2011) with the solar luminosity constraint, to
fix the principal pp-chain fluxes.  That is, the sum of the  rates for the ppI, ppII, and ppIII cycles,
weighted by the energy each cycle deposits in the Sun, is fixed by the photon luminosity.  
Consequently the ``solar values" are
not strictly measured ones, but derived assuming a steady-state Sun.

Table \ref{tab:fluxes} also gives fluxes for two solar models, reflecting the metallicity uncertainties
mentioned previously.  The first model
uses abundances for volatile elements that were obtained from an absorption line analysis in which the photosphere
was treated in 1D,  yielding (Z/X)$_S=0.0229$ \citep{GS98}.  As Solar Fusion II 
cross sections are used as well, this model is denoted GS98-SFII.  The second, denoted AGSS09-SFII, 
takes abundances from a 3D photospheric model, yielding 
(Z/X)$_S$=0.0178 \citep{AGSS09}.  The solar core is sensitive
to metallicity, as free-bound/bound-free transitions in metals are an important contributor
to the opacity.  A low-metallicity Sun, as in model AGSS09-SFII, produces a somewhat cooler core (by $\sim$ 1\%),
and thus lower fluxes of temperature-sensitive neutrinos, such as those from $^8$B $\beta$ decay.

The SSM fluxes for the CN I cycle $\beta$ decays of $^{13}$N and $^{15}$O
and the CNO II cycle decay of ${}^{17}$F are also shown.  
Despite the minor role the CN cycle
plays in solar energy generation, these fluxes are significant.  The excess in the flux of $^{13}$N neutrinos 
relative to $^{15}$O neutrinos in Table \ref{tab:fluxes} is a consequence of the out-of-equilibrium burning
of the CN cycle discussed previously.

The SSM uncertainties given in Table \ref{fig:cycles} are generated from the uncertainties
assigned to approximately 20 model input parameters, denoted $\beta_j$.  These include the solar
age, present-day luminosity, opacities, the diffusion constant, the S-factors for the
pp chain and CN cycle, and the various metal abundances (key elements such as
C, N, O, Ne, Mg, Si, S, Ar, and Fe). The consequences
of input SSM uncertainties on observables are typically parameterized through logarithmic partial derivatives $\alpha(i,j)$,
determined by calculating the SSM response to  variations
in the $\beta_j$.  For example, in the case of the neutrino fluxes $\phi_i$, the
\begin{equation}
\alpha(i,j)  \equiv  {\partial   \ln{\left[  \phi_i/\phi_i(0)  \right]}  \over
\partial \ln{\left[ \beta_j / \beta_j(0)\right]}}
\end{equation}
can be found in the SSM updates of
\citet{PGS2008} and \citet{Serenellialpha}.  Here
$\phi_i(0)$  and  $\beta_j(0)$   denote  the  SSM  best  values.  

The partial derivatives
define  the power-law  dependencies of
neutrino fluxes with respect to the SSM best-value prediction $\phi_i(0)$,
\begin{equation}
\phi_i   \sim   \phi_i(0)  \prod_{j=1}^{19}   \left[   {\beta_j  \over   \beta_j(0)}
\right]^{\alpha(i,j)}=\phi_i(0)  \prod_{j=1}^{19}   \left[ 1+  \delta \beta_j
\right]^{\alpha(i,j)},
\label{eq:prod}
\end{equation}
where $\delta \beta_j \equiv \Delta \beta_j/\beta_j(0)$ is the
fractional uncertainty of input parameter $\beta_j$ with respect to its
SSM best value. As this expression 
separates the impact of SSM parameter variations on $\phi_i$ into a
solar piece -- the infinitesimal SSM response described by $\alpha(i,j)$ --
and a laboratory or theory piece -- the estimated uncertainty $\delta \beta_j$ of 
an input parameter, the effects of parameter variations can be explored
without repeating SSM calculations.

\begin{table}
\caption{SSM neutrino fluxes from the GS98-SFII and AGSS09-SFII SSMs, with
associated uncertainties (averaging over asymmetric uncertainties).  The solar values
come from a luminosity-constrained analysis of all available data by the Borexino Collaboration.}
\vspace{0.4cm}
\label{tab:fluxes}
\scalebox{0.8}{
\begin{tabular}{lccccc}
\hline \hline
 $\nu$ flux & E$_\nu^\mathrm{max}$ (MeV) & GS98-SFII & AGSS09-SFII & Solar & units \\
\hline
p+p$\rightarrow^2$H+e$^+$+$\nu$ & 0.42 & $5.98(1 \pm 0.006)$ & $6.03(1 \pm 0.006)$ & $6.05(1^{+0.003}_{-0.011})$ & 10$^{10}$/cm$^2$s \\
p+e$^-$+p$\rightarrow^2$H+$\nu$ & 1.44 & $1.44(1 \pm 0.012)$ & $1.47(1 \pm 0.012)$ & $1.46(1^{+0.010}_{-0.014})$ & 10$^8$/cm$^2$s\\
$^7$Be+e$^-$$\rightarrow^7$Li+$\nu$ & 0.86 (90\%) & $5.00(1 \pm 0.07)$ & $4.56(1 \pm 0.07)$ & $4.82(1^{+0.05}_{-0.04})$ & 10$^9$/cm$^2$s\\
 & 0.38 (10\%) & & & & \\
$^8$B$\rightarrow^8$Be+e$^+$+$\nu$ & $\sim$ 15 & $5.58(1 \pm 0.14)$ & $4.59(1 \pm 0.14)$ & $5.00(1\pm 0.03)$ & 10$^6$/cm$^2$s\\ 
${}^3$He+p$\rightarrow^4$He+e$^+$+$\nu$  & 18.77 & $8.04(1 \pm 0.30)$ & $8.31(1 \pm 0.30)$ & --- & 10$^3$/cm$^2$s\\
$^{13}$N$\rightarrow^{13}$C+e$^+$+$\nu$  & 1.20 & $2.96(1 \pm 0.14)$ & $2.17(1 \pm 0.14)$ &$\leq 6.7$ & 10$^8$/cm$^2$s\\ 
$^{15}$O$\rightarrow^{15}$N+e$^+$+$\nu$  & 1.73 & $2.23(1 \pm 0.15)$ & $1.56(1 \pm 0.15) $ &$\leq 3.2$ & 10$^8$/cm$^2$s\\ 
$^{17}$F$\rightarrow^{17}$0+e$^+$+$\nu$  & 1.74 & $5.52(1 \pm 0.17)$ & $3.40(1 \pm 0.16) $ &$\leq 59.$ & 10$^6$/cm$^2$s\\ 
$\chi^2/P^\mathrm{agr}$ & & 3.5/90\% & 3.4/90\% & & \\
\hline \hline
\end{tabular}
}
\end{table}

The solar abundance problem is characterized by large systematic differences in the SSM $\beta_j$
for key abundances.  Consequently the differences in the GS98-SFII and AGSS09-SFII
SSM neutrino flux predictions of Table \ref{tab:fluxes} exceed their respective internal statistical uncertainties, in several cases.
The Table summarizes key helioseismic predictions of both models, with the low-Z AGSS09-SFII SSM
predictions being in significantly poorer agreement with the data.  

\begin{figure}
\includegraphics[width=13cm]{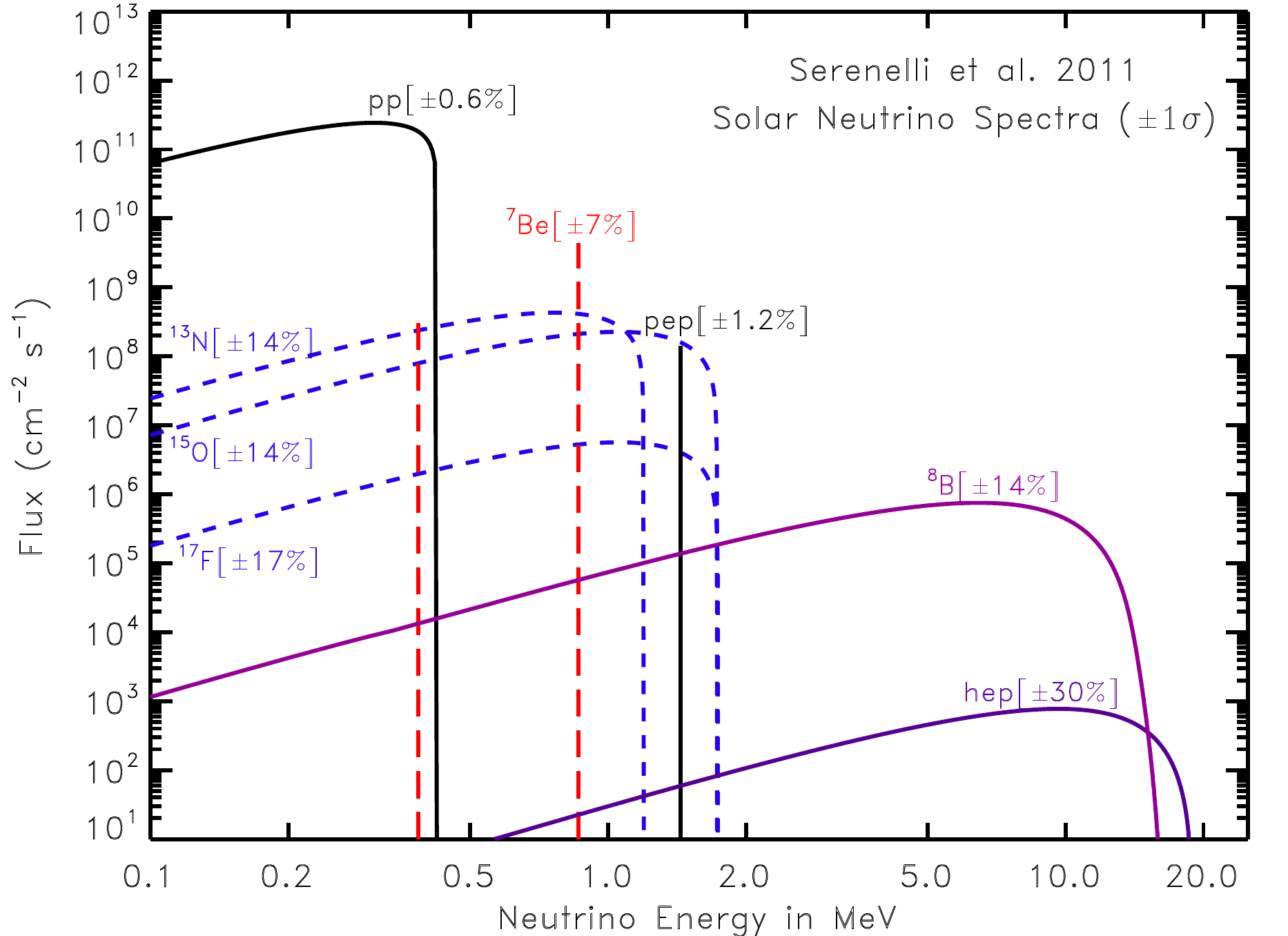}
\caption{(Color online) The solar neutrino spectrum, along
with the SSM uncertainties \citep{Serenelli11}.  A weak branch from the 
$\beta$ decay of ${}^{17}$F that contributes from the CN II cycle is included. The units for the continuous
sources are cm$^{-2}$ s$^{-1}$MeV$^{-1}$.}
\label{fig:fluxes}
\end{figure}

The spectra of solar neutrinos emitted by the Sun is shown in Fig. \ref{fig:fluxes}.  This familiar figure
its somewhat idealized: it includes competing $\beta$ decay and EC branches for the 
p+p reaction,  but omits the EC lines that accompany the other $\beta$ decay reactions of Table \ref{tab:fluxes}.
The EC branching ratio increases with increasing Z and decreasing Q-value.
Thus, among the omitted cases, EC is most significant for the CNO cycle reactions, where
rate estimates were made by \cite{SFR}.  The EC lines shown in
the figure are in fact thermally broadened, as they occur in a plasma.  Finally, at energies $\lesssim$ 10 keV 
below the scale of Fig. \ref{fig:fluxes}, there is a contribution from
neutrino pairs of all flavors produced thermally in the solar core \citep{HW2000}: while the flux of these
neutrinos is modest, the peak flux density of
$\sim$ 10$^9$/cm$^2$/s/MeV exceeds that of all solar $\beta$ decay sources, except for the pp neutrinos.

In the following sections, measurements of the various sources can provide unique information on the
structure and composition of the Sun, and on the properties of neutrinos, including how those
properties depend on the surrounding environment.

\section{EXPERIMENTS:  NEUTRINOS AND HELIOSEISMOLOGY}
\label{sec:three}
The SSM is a model of the Sun's interior structure, rather than the more complicated behavior of the
convective envelope.  The two main tools by which we can probe the solar interior and thus test
the validity of the SSM  are neutrino spectroscopy and helioseismology.  Neutrino fluxes are
sensitive to core temperature -- a property that responds to changes in the opacity and composition --
and to nuclear cross sections.  Helioseismic maps of the sound speed throughout much
of the solar interior have achieved accuracy of a few parts in 1000, constraining   the solar density and
 pressure profiles, and determining  
rather precisely the boundary between the Sun's radiative core and
convective envelope.

\subsection{Neutrino Spectroscopy: Early Experiments}
\subsubsection{The Chlorine Experiment}
Radiochemical detection of neutrinos by $^{37}$Cl($\nu_e$,e$^-$)$^{37}$Ar was suggested by \cite{Pontecorvo46}
and explored in more detail by \cite{Alvarez49}, who was interested in the possibility
of a reactor neutrino experiment to detect a Majorana neutrino.  
Ray Davis's efforts to develop a practical detector began with
his Brookhaven experiment \citep{Davis55},  which used a 1000-gallon tank of perchloroethylene (C$_2$Cl$_4$)
placed 19 feet underground.  This yielded an upper bound on the rate for
solar neutrino reactions of $\sim$ 40,000 SNU.
Subsequent developments are described by
\cite{Lande10}.  Construction began on
the Homestake detector in 1965, with first results announced in 1968, and with measurements
continuing until 2002, when the Homestake Mine closed.  The final result \begin{equation}
\langle \sigma \phi \rangle = 2.56 \pm 0.16 \pm 0.16 \mathrm{~SNU}
\end{equation}
is about a factor of three below historical SSM best values.  (The SSM GS98-SFII rate is 8.00 $\pm$ 0.97 SNU.)

The experiment exploited fortuitous properties of ${}^{37}$Ar in achieving nearly
single-atom counting.  The average solar neutrino reaction rate
in the tank was 0.48 counts/day, above a background dominated by
cosmogenics of 0.09 counts/day.  As a noble gas that does not interact chemically, argon can
be extracted with high efficiency ($\gtrsim$ 95\%) from large volumes of organic liquid.  The ${}^{37}$Ar half life
of $\sim$ 35 days allowed tank concentrations to build up
over a saturation time of $\sim$ two months, yet also made ${}^{37}$Ar counting via electron capture feasible.
As the decay populates the ground state of ${}^{37}$Cl, the signal is the 2.82 keV Auger
electron produced as the electrons in ${}^{37}$Cl adjust to fill the K-shell vacancy.
Davis developed
miniaturized gas proportional counters for this counting.
Taking into account detector efficiencies and losses due to ${}^{37}$Ar decaying in the tank,
the number of Ar atoms counted per year was $\sim$ 25.

The chlorine experiment was primarily sensitive to the temperature-dependent neutrino
fluxes produced in the ppIII and ppII cycles (${}^8$B $\sim$ 75\%, ${}^7$Be $\sim$ 16\%).
For this reason the source of the ``solar neutrino problem" was not immediately apparent.
Many candidate explanations were offered over the next 30 years, with many of these proposing
changes in the SSM to produce a somewhat cooler core.

\subsubsection{Kamiokande II/III}
Confirmation of the ${}^{37}$Cl results came 21 years later, from a 
detector originally designed for proton-decay studies.  The Kamiokande I
proton decay detector was upgraded in the early 1980s to Kamiokande II/III, a
3.0-kiloton imaging water Cherenkov detector capable of detecting solar and other
low-energy neutrinos.  The neutrino signal is the Cherenkov light emitted by recoiling
electrons after elastic scattering (ES),
$\nu_x + e^- \rightarrow \nu_x^\prime + e^-$, which
is sensitive to both electron and heavy-flavor neutrinos, though with the differential
sensitivity $\sigma(\nu_e)/\sigma(\nu_\mu) \sim$ 6.   For incident neutrino energies $\gg m_ec^2$,
the electron is scattered in the forward
direction.  Thus, by correlating event directions with the position of the Sun,
one can cut away a large background uncorrelated with solar position,
to reveal solar neutrino events in a forward cone.  

The inner 2.14 kilotons of the detector was viewed by 948 20-inch Hamamatsu photomultiplier (PMT) detectors,
providing $\sim$ 20\% coverage.   The innermost 0.68 kilotons of the detector
served as the fiducial volume for solar neutrino detection.
Kamiokande II operated with a $\sim$ 9 MeV threshold, 
which was later reduced to 7.5 and 7.0 MeV in Kamiokande III. 

The improvements made in Kamiokande II to enable solar neutrino detection included
water purification to reduce low-energy backgrounds associated with radon
and uranium as well as electronics upgrades to improve timing, vital for
the reconstruction of the interaction vertices and directions of low-energy electrons,
and thus in more cleanly defining a fiducial volume for solar neutrino events.
After water-sealing the cavity holding the main detector, 
the outer portion of the detector was instrumented with 123 PMTs
to serve as a muon anti-counter, and
additional water
was added to shield against $\gamma$s from the surrounding rock.   Kamiokande III
included improvements in the electronics, water purification system,
event reconstruction and selection tools, and Monte Carlo detector simulations software.

The first production run of Kamiokande II began in January 1987.  The detection of
$^8$B solar neutrinos based on 450 days of data extending through May 1988
was announced by \cite{KamiokandeA}.  The measured flux of $^8$B neutrinos
with energies above 9.3 MeV was found to be $0.46 \pm 0.13(\mathrm{stat}) \pm 0.08 (\mathrm{syst})$
of the SSM value, confirming the deficit seen by Davis and collaborators. 
Kamioka II/III ran until February 1995, collecting 2079 days of data.
The combined analysis of all data sets yielded \citep{KamiokandeB}
\begin{equation}
\phi(^8\mathrm{B}) = (2.80 \pm 0.19 (\mathrm{stat}) \pm 0.33 (\mathrm{sys})) \times 10^6/\mathrm{cm}^2 \mathrm{s},
\end{equation}
or 50\% (61\%) of the GS98-SFII (AGSS09-SFII) SSM result.

The Kamiokande II/III detector was the first to record solar neutrinos event by event, 
establish their solar origin through correlation with the direction to the Sun, and provide
direct information on the $^8$B spectrum through the recoil electron spectrum from ES.

\subsubsection{The Gallium Experiments}
Two radiochemical gallium experiments exploiting the reaction $^{71}$Ga($\nu_e$,e$^-$)$^{71}$Ge,
SAGE and GALLEX/GNO, began solar neutrino measurements in January 1990 and May 1991,
respectively.  SAGE, which continues to operate, uses a target of 50 tons of Ga metal, heated
to 30$^\circ$C so that the metal remains molten, and has reported results for 168 extractions through
December 2007.  The experiment is located in the Baksan Neutrino Observatory, under Mt.
Andyrchi in the Caucasus.  GALLEX, which used 30 tons of Ga in the form of a GaCl$_3$ solution,
operated through 1997, and its successor GNO continued through 2003.  GALLEX and GNO were 
mounted in the Gran Sasso National Laboratory, near L'Aquila, Italy.

Gallium, first proposed as a solar neutrino detector by \cite{kuzmin}, has a low threshold
for solar neutrino absorption (233 keV) and a strong Gamow-Teller transition to the ground state of $^{71}$Ge. 
This leads to a large cross section for absorbing the low-energy pp neutrinos. 
As $^{71}$Ge has a half life of 11.43 days, a radiochemical experiment analogous to that done
for chlorine was proposed,  though the chemistry of Ge recovery is considerably
more complicated than that for Ar.  Because of its
pp neutrino sensitivity, the Ga experiment has
a minimum astronomical counting rate of 79 SNU, assuming
only a steady-state Sun and standard-model (SM) weak interaction physics.  That is, any combination
of ppI, ppII, and ppIII burning consistent with the Sun's observed luminosity will lead to a solar neutrino capture rate 
at or above this value \citep{Book}.  Thus the gallium experiment had the potential
to yield a result that would require a ``new physics" solution to the solar neutrino problem.

In 1974 Davis and collaborators
began work on the chemistry of Ge recovery for both GaCl$_3$ solution and Ga metal,
conducting a 1.3-ton pilot experiment using GaCl$_3$ in 1980-82 to demonstrate the procedures
later used by GALLEX.  The method recovers Ge
as GeCl$_4$ by bubbling nitrogen through the solution, then scrubbing the gas.
The Ge can be further concentrated and purified, converted into GeH$_4$, then counted in
miniaturized gas proportional counters similar to those used in the chlorine experiment.

In the SAGE experiment with room-temperature liquid metal, the ${}^{71}$Ge is separated
by mixing into the gallium a solution of hydrogen peroxide and dilute hydrochloric acid,
which produces an emulsion, with the germanium migrating to the surface of the emulsion as
droplets, where it can be oxidized and dissolved by hydrochloric acid.  The Ge is extracted as GeCl$_4$,
purified and concentrated, synthesized into GeH$_4$, then counted as in the GALLEX experiment.
In both GALLEX and SAGE, the overall efficiency of the chemical procedures can be determined
by introducing Ge carrier.

A unique aspect of the gallium experiments was the neutrino source experiments
done to check overall experimental procedures --
chemical extraction, counting, and analysis techniques.  The calibrations also checked the capture cross section, 
as two excited-state transitions not constrained by the ${}^{71}$Ge lifetime contribute to $^7$Be neutrino capture.   Two GALLEX
calibrations and the first SAGE calibration were done with $^{51}$Cr sources, while the second
SAGE calibration used an $^{37}$Ar source.  Source intensities were $\sim$
0.5 MCi.  The weighted average \citep{abdurashitov09} for the four calibrations, expressed as the ratio $R$ of measured ${}^{71}$Ge
to that expected due to the source strength, is $R=0.87 \pm 0.05~ (1 \sigma)$.  
The discrepancy, which exceeds two standard deviations, has attracted some attention due to
other short-baseline neutrino anomalies \citep{gavrin2011,sterile}.

SAGE began taking data in December 1989.  The capture rate limit obtained from 
five extractions with 30 tons of gallium,
$\lesssim$ 79 SNU (90\% c.l.)  \citep{SAGE1}, coincided with the minimum astronomical
value.   The most recent SAGE combined analysis for all 168 extractions yielded \citep{abdurashitov09}
\begin{equation}
\langle \sigma \phi \rangle_{SAGE} = 65.4^{+3.1}_{-3.0} \mathrm{(stat)}^{+2.6}_{-2.8} \mathrm{(syst)}~\mathrm{SNU},
\end{equation}
or approximately half the un-oscillated SSM best value.  GALLEX began taking data in May 1991 and announced
first results a year later, a counting rate based on 14 extractions of 83 $\pm$ 19 (stat) $\pm$ 8 (syst) SNU \citep{gallex1}.
GALLEX completed four campaigns, I through IV, running until 1997.  From 65 extractions
and 1594 days of data, and including updates due to new pulse-shape analysis methods, GALLEX found \citep{gallex4,gallex4a}
\begin{equation}
\langle \sigma \phi \rangle_{GALLEX~I-IV} = 73.1^{+6.1}_{-6.0} \mathrm{(stat)}^{+3.7}_{-4.1} \mathrm{(syst)}~\mathrm{SNU}.
\end{equation}
A number of improvements in Ge extraction procedures, electronics, counter efficiency calibrations, and radon
event characterization were incorporated into the follow-up experiment GNO.  The experiment accumulated 
1687 days of running between May 1998 and April 2003.  The
counting rate from the 58 extractions is
\begin{equation}
\langle \sigma \phi \rangle_{GNO} = 62.9^{+5.5}_{-5.3} \mathrm{(stat)}^{+2.5}_{-2.5}  \mathrm{(syst)}~\mathrm{SNU}.
\end{equation}
The weighted average of SAGE, GALLEX, and GNO results is
\begin{equation}
\langle \sigma \phi \rangle_{SAGE+GALLEX+GNO} = 66.1 \pm 3.1~\mathrm{SNU},
\end{equation}
with all
uncertainties combined in quadrature \citep{abdurashitov09}. The SSM GS98-SFII rate is
126.6 $\pm$ 4.2 SNU.

\subsubsection{Hints of New Physics}
\begin{figure}
\includegraphics[width=13cm]{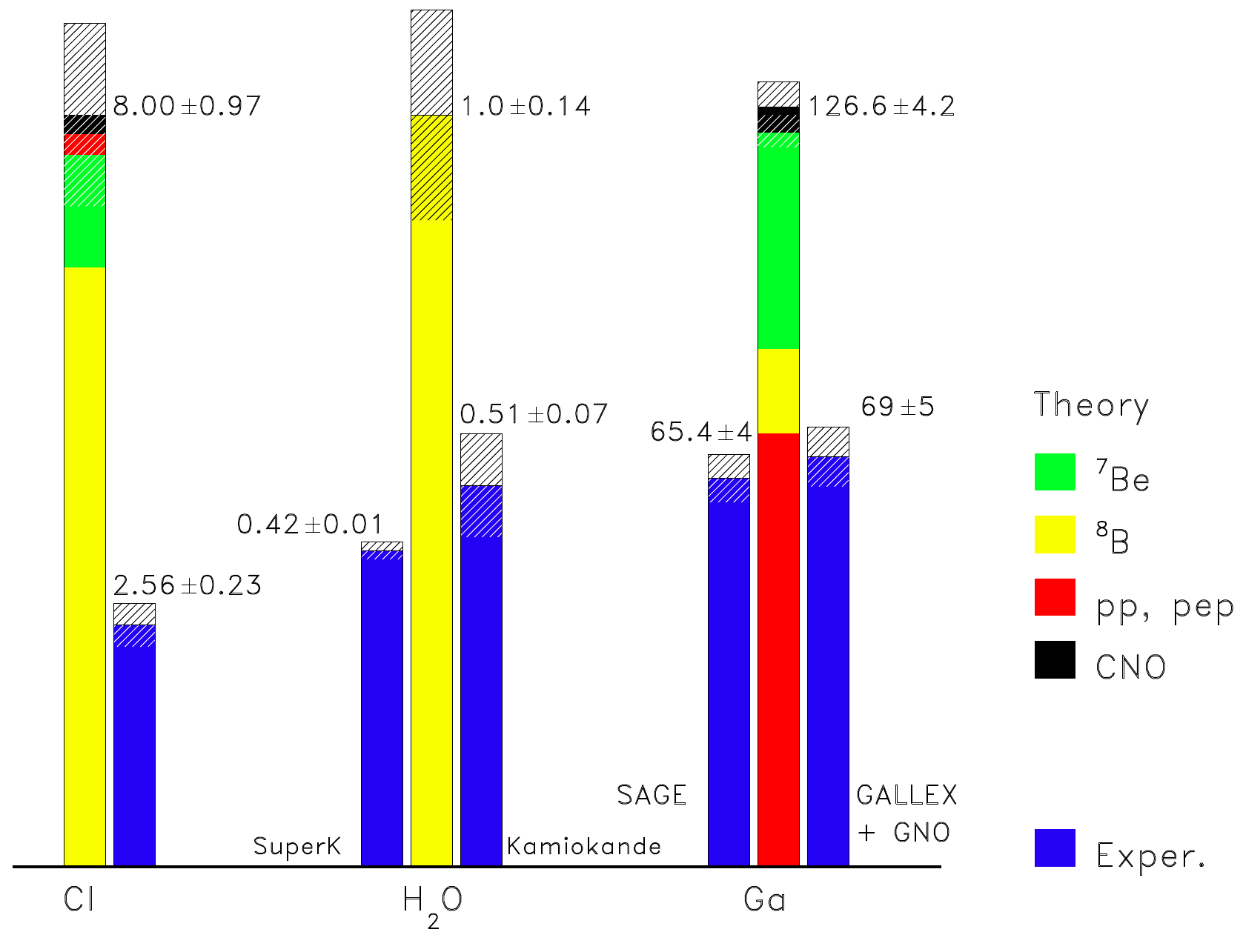}
\caption{(Color online) Comparison of the measured neutrino rates for the chlorine,
Kamioka II/III, and SAGE/GALLEX/GNO experiments with the contemporary
SSM GS98-SFII, assuming unoscillated fluxes \citep{Serenelli11}.}
\label{fig:old}
\end{figure}

In Fig. \ref{fig:old} results of the early experiments are compared to the predictions of the contemporary GS98-SFII SSM.  Not only are the results in
disagreement with the SSM, but the pattern of discrepancies is not easily
reproduced even if one entertains the possibility of substantial
variations in that model.
By the early 1990s several analyses \citep{White93,Hata94,Parke95} had pointed
to apparent contradictions in the pattern of fluxes with respect to SSM predictions,
\begin{equation}
\label{eq:fluxpattern}
\phi(\mathrm{pp}) \sim 0.9 \phi^\mathrm{SSM}(\mathrm{pp}) ~~~~~
\phi(^7\mathrm{Be}) \sim 0~~~~~
\phi(^8\mathrm{B}) \sim 0.4 \phi^\mathrm{SSM}(^8\mathrm{B}).
\end{equation} 
Now variations in the SSM affect the neutrino fluxes principally through their impact on the core temperature
$T_C$.  As
\begin{equation}
{\phi(^8\mathrm{B}) \over \phi(\mathrm{pp})} \sim T_C^{22},
\end{equation}
the observation that $\phi(^8\mathrm{B})/\phi(\mathrm{pp}) \sim 0.4 \phi^\mathrm{SSM}(^8\mathrm{B})/ \phi^\mathrm{SSM}(\mathrm{pp})$
would seem to require a cooler solar core, $T_C \sim 0.95~T_C^\mathrm{SSM}$.  On the other hand, as
\begin{equation}
{\phi(^7\mathrm{Be}) \over \phi(^8\mathrm{B})} \sim T_C^{-12},
\end{equation}
the observation that $\phi(^7\mathrm{Be})/\phi(^8\mathrm{B}) << \phi^\mathrm{SSM}(^7\mathrm{Be})/ \phi^\mathrm{SSM}(^8\mathrm{B})$
would seem to require a hotter core, $T_C >T_C^\mathrm{SSM}$, a contradiction.   
Model-independent analyses 
assuming undistorted neutrino spectra and a steady-state Sun (so that neutrino fluxes are constrained by the
Sun's luminosity) were done by \cite{HBL} and \cite{HR}.  Their calculations showed that the probability
of solutions without new particle physics were in a range of  $\sim$ 2-3\%.   \cite{HR} further argued that if the luminosity
constraint were relaxed, this probability would still be limited to $\sim$ 4\%.   The likelihood of a new-physics 
solution to the solar neutrino problem was high.

The conclusion -- that the solar neutrino problem might have its origin outside of astrophysics -- was 
additionally supported by a growing body of evidence from helioseismology that
validated SSM descriptions of the Sun's interior structure.

\subsection{Helioseismology}
Measurements and analysis of Doppler shifts of photospheric absorption lines show that the Sun's surface
oscillates with amplitudes $\sim$ 30m and velocities $\sim$ 0.1 m/s, reflecting a variety of interior modes \citep{Gizon02}.
Turbulence within the Sun's convective zone acts as a random driver of sound waves propagating through
the gas.  Specific frequencies are enhanced as standing waves, normal eigenmodes that reflect details
of solar structure.  Here we summarize the basics of solar oscillations, referring readers to \cite{Chaplin13}
for a more detailed discussion.

The SSM is characterized by quasi-static pressure $p(r)$, density $\rho(r)$, 
temperature $T(r)$,
entropy $s(r)$, gravitational potential $\phi(r)$, and nuclear energy generation $
\epsilon(r)$ profiles that
are functions of the radial coordinate $r$. 
One can perturb the SSM by introducing
small displacements $\delta \vec{r}$ and associated velocities $v(\vec{r})= \partial \delta \vec{r}/ \partial t$,  
then seek small-amplitude normal-mode solutions \citep{CD02}
\begin{equation}
\rho(\vec{r},t) \equiv \rho_0(r) + \rho^\prime(\vec{r},t)~~~~~~~~\rho^\prime(\vec{r},t)\sim \rho^\prime(r) ~Y_{lm}(\theta,\phi) e^{i \omega t}
\end{equation}
that might account for observed solar surface oscillations. 

Solar oscillations can be treated in the adiabatic approximation because the timescale for heat exchange
is much longer than the oscillation periods of interest.  From the adiabatic index $\Gamma_1$ describing the power-law dependence of the
pressure on the density and the associated sound speed $c(r)$,
\begin{equation}
\Gamma_1 \equiv \left({\partial~ \mathrm{log}~ p(r) \over \partial ~\mathrm{log}~ \rho(r)} \right)_s~~~~~~~p(r) = {1 \over \Gamma_1} \rho(r) c^2(r),
\end{equation}
one can define an auxiliary field $\Psi(\vec{r})= c^2 \sqrt{\rho(r)} ~\vec{\nabla} \cdot \delta \vec{r}$.   In the Cowling (neglecting perturbations to the gravitational field) and adiabatic approximations \citep{Gough84}
\begin{equation}
{d^2 \Psi_l(r) \over dr^2} + {1 \over c^2} \left[ \omega^2 - \omega_\mathrm{co}^2-{l(l+1)c^2 \over r^2} \left(1-{N^2 \over \omega^2} \right) \right] \Psi_l(r) \equiv
\left( {d^2 \over dr^2} + {\omega^2_\mathrm{eff} \over c^2}\right)  \Psi_l(r) \sim 0,
\label{eq:helio}
\end{equation}
where propagating (evanescent) solutions exist for $\omega_\mathrm{eff}^2 > 0$ ($<$0).
This eigenvalue problem is
governed by the buoyancy, or Brunt-V\"{a}is\"{a}l\"{a}, frequency $N(r)$,
\begin{equation}
N^2(r) = { G m(r) \over r} \left( {1 \over \Gamma_1} {d\mathrm{log}~ p(r) \over dr} - {d\mathrm{log} ~\rho(r) \over dr} \right),
\end{equation}
which turns negative in the convective zone but is positive and roughly constant in the radiative interior;
the Lamb frequency,
\begin{equation}
S_l^2(r) = {l(l+1) c^2 \over r^2},
\end{equation}
which diverges for $r \rightarrow 0$ if $l > 0$; and the acoustic cutoff frequency, 
which depends on the density scale height $H(r)$ and sound speed,
\begin{equation}
\omega_\mathrm{co}(r) = {c \over 2H} \sqrt{1-2 {d H \over dr}} 
\mathrm{~~~where~~~}
H(r) \equiv - \left({1 \over \rho(r)} {d \rho(r) \over dr} \right)^{-1},
\end{equation}
and determines the outer turning point where $\omega \sim \omega_{co}$.  Eigensolutions of Eq. (\ref{eq:helio})
can be found for discrete frequencies $\{\omega_{nl} \}$, where $n$ is the radial order: there is no dependence 
on $m$ because all azimuthal modes for fixed $n,l$ are degenerate by spherical symmetry.
The assumptions leading to Eq. (\ref{eq:helio}) can be justified except when $n$ and $l$ are small,
or when $l \ll n$ \citep{Gough84}.

As $ \omega \gg \omega_{co}$ everywhere except near the surface, the solar
regions supporting propagating solutions are determined by
\begin{equation}
\omega^2_\mathrm{eff} \sim \omega^2 -{l(l+1) c^2 \over r^2} \left( 1 - {N^2 \over \omega^2} \right) > 0.
\label{eq:eff}
\end{equation}
Two different families of solutions exist.  The g-mode family is determined by the conditions $\omega^2 \ll N^2$ and
$\omega^2 \ll S_l^2$.  Consequently g-mode propagation is confined to the solar radiative interior.
The second family, the acoustic oscillations or p-modes, are the modes that have been observed in the Sun.
If $\omega^2 \gg N^2$, then Eq. (\ref{eq:eff}) and the requirement $\omega^2_\mathrm{eff}>0$ define the inner turning-point radius
\begin{equation}
r_\mathrm{turning} \sim {c(r) \over \omega} \sqrt{l(l+1)} .
\end{equation}
Qualitatively it is clear that the dependence of the eigenfrequencies on $l$ can provide localized sensitivity to
$c(r)$, with modes of low $l$ penetrating more deeply into the solar interior.  Because the eigenfrequencies depend on $c(r)$,
the p-mode observations constraint the solar pressure and density profiles.

Similar radial sensitivity is found for the g-modes.  The condition $\omega \ll N(r)$ allows propagation in the deep interior,
as Eq. (\ref{eq:eff}) guarantees that $\omega^2_\mathrm{eff} >0$ for $r$ sufficiently small.   While in principle this
suggests sensitivity to $c(r)$ in the solar core, g-modes are damped in the convective envelope, making
observation difficult.  No undisputed detection exists to date \citep{Appourchaux10}.

\begin{figure}
\includegraphics[width=13cm]{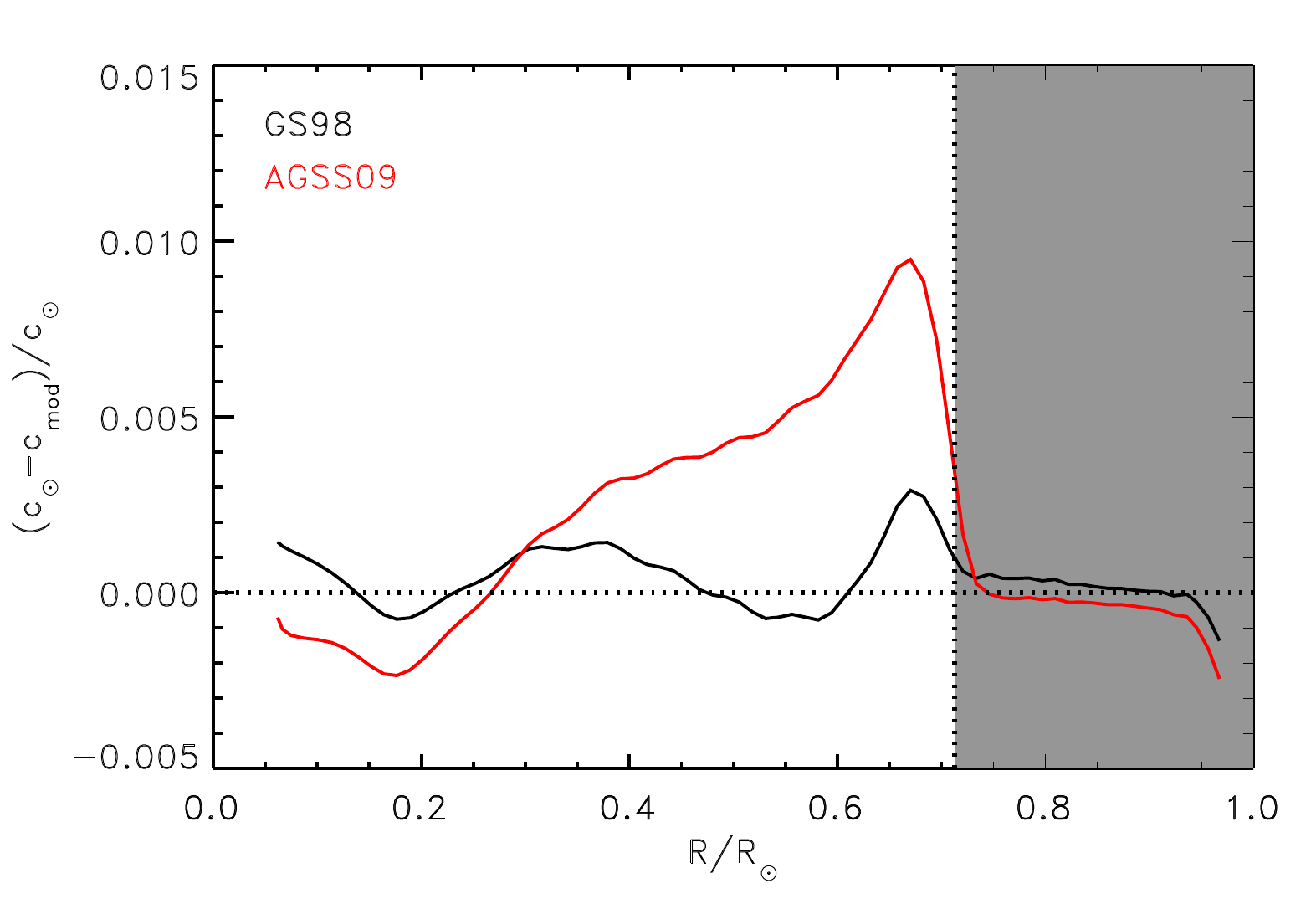}
\caption{(Color online) The relative sound speed $(c(r)_\mathrm{solar}-c(r)_\mathrm{SSM})/c(r)_\mathrm{SSM}$
where $c(r)_\mathrm{SSM}$ is the SSM result and $c(r)_\mathrm{solar}$ the solar profile extracted 
from BiSON data.  The black and red profiles correspond to the high-metallicity GS98-SFII and
low-metallicity AGSS09-SFII SSMs, respectively.}
\label{fig:helio}
\end{figure}

The significant effort invested in helioseismological measurements and analysis has yielded a
rather precise map of $c(r)$ over the outer 90\% of the Sun by radius.  The solar profile used in Fig. \ref{fig:helio}
was obtained by \cite{Basu09} from an analysis that included 4752 days of BiSON data
(http://bison.ph.bham.ac.uk).
The comparison SSMs are AGSS09-SFII ((Z/X)$_S=0.0178$)
and GS98-SFII ((Z/X)$_S=0.0229$).   

GS98-SFII is representative of models that were in
use in the 1990s: the generally good agreement with the solar $c(r)$ ($\sim$ 0.2\% apart from a narrow region just
below the convective boundary) was taken as strong
support for the SSM, helping to reinforce the conclusion
that the solar neutrino problem might have a particle physics origin.  
Helioseismic data forced improvements in the SSM, such as the inclusion 
of helium and heavy-element diffusion \citep{BP92,BP95}.  The suggestion from early 
solar neutrino experiments and helioseismology that new particle physics could be the source of
the solar neutrino problem provided additional motivation for
a new generation of sophisticated experiments with high statistics and
sensitivity to neutrino flavor, spectral distortions, and day-night differences, described below.

\subsection{Super-Kamiokande}
Super-Kamiokande, the successor to the Kamiokande detector, is a 50-kton cylindrical water Cherenkov detector located in the Kamioka Mine at a depth of $\sim$ 2.03 km of water (flat site equivalent).   The inner 32-ktons of water is viewed by $\sim$ 11,100 20" photomultipliers (40\% coverage),
with 22.5-ktons serving as the fiducial volume for detecting solar neutrinos.  The detector has operated
at (total) electron energy thresholds for solar neutrinos ranging from 7.0 to the present 4.0 MeV, so
that detection is limited to $^8$B and hep neutrinos.    While in ES
the energy of the incident neutrino is shared between the scattered electron and outgoing neutrino, electron detection
provides some sensitivity to the initial neutrino spectrum and thus to distortions associated with neutrino
oscillations  \citep{SKweb,SKNIM}.  The electron energy resolution at 10 MeV
is $\sim$ 16\%.

The detector began operations in 1996, progressing from phase I to the current phase IV.  
Super-Kamiokande I recorded neutrino events for approximately
five years, determining an $^8$B neutrino flux of $\phi(^8\mathrm{B})$ = (2.35 $\pm$ 0.02 (stat) $\pm$ 0.08 (syst)) $\times$ 10$^6$/cm$^2$/s
from events recorded above 5 MeV, assuming an undistorted spectrum \citep{SKI}.  The measured rate
variation of $\sim$ 7\%
over the year is consistent in magnitude and phase with the effects of the Earth's orbital eccentricity, 1.7\%.
No evidence was found for spectral distortions or day-night differences, two signatures of
neutrino oscillations in matter.

The detector was drained following phase I for repairs and maintenance.  During refilling the implosion
of a phototube led to a blast wave that destroyed most of the lower part of the detector.  Super-Kamiokande II
was subsequently rebuilt with the remaining phototubes enclosed in blast shields.  Despite the reduced phototube
coverage of 19\% and resulting higher threshold of 7 MeV, Super-Kamiokande II ran successfully
as a solar neutrino detector for three years, beginning in late 2002.  The deduced rate,
$\phi(^8\mathrm{B})$ = (2.38 $\pm$ 0.05 (stat) $^{+0.16}_{-0.15}$ (sys)) $\times$ 10$^6$/cm$^2$/s, is consistent
with Super-Kamiokande I results.  No spectral distortion was detected, and the day-night difference
was again consistent with zero at $1\sigma$ \citep{SKII}.

Super-Kamiokande III \citep{SKIII} collected nearly two years of data between October 2006 and August 2008,
operating with a fully restored set of 11129 PMTs equipped with blast shields, providing 40\% phototube
coverage.  Improvements made to the water purification system, event reconstruction and selection
tools, and the Monte Carlo detector simulation software resulted in a reduced systematic uncertainty of $\pm 2.1$\%.
The observed event rate for electrons between 5.0 and 20 MeV is equivalent to an unoscillated $^8$B neutrino
flux of (2.39 $\pm$ 0.04 (stat) $\pm 0.05$ (sys)) $\times$ 10$^6$/cm$^2$/s \citep{Smy}.  No significant spectral
distortion was observed. 

Preliminary results from 1069 days of running for Super-Kamiokande IV were reported at Neutrino 2012 \citep{Smy}.
This latest phase of Super-Kamiokande includes new electronics, an improved Monte Carlo model of the
trigger efficiency,  higher efficiency due to relaxed cuts against radioactivity backgrounds, and a lower
threshold of 4 MeV (total energy).  

\subsection{Sudbury Neutrino Observatory}
A second remarkable detector,  the Sudbury Neutrino Observatory (SNO), was constructed two
kilometers below ground, within the INCO Creighton nickel mine, Ontario,
Canada \citep{SNOreview}.  A  kiloton of heavy water was contained in a 12m-diameter spherical
acrylic vessel. A surrounding array of 9500 20-cm PMTs viewed the inner
volume, providing $\sim$ 56\% coverage.  Seven
kilotons of light water provided a 5m buffer between the central detector and the surrounding rock walls.

SNO was proposed by \cite{HChen}, who recognized
the advantages of the multiple detection channels
that could be introduced by replacing the hydrogen
in an ordinary water Cherenkov detector with deuterium.  The flux of higher energy solar electron neutrinos 
can be probed with the charged-current  (CC) reaction
\begin{equation}
\nu_e + d \rightarrow p + p + e^-,
\end{equation}
with detection of the produced electron.
As the Gamow-Teller strength is concentrated
near the 1.44 MeV breakup threshold for deuterium, the electrons carry off most of the energy,
and thus provide significant information
on the incident neutrino spectrum.
A second channel, the neutral-current (NC) reaction
\begin{equation}
\nu_x + d \rightarrow \nu_x^\prime + n + p,
\end{equation}
is independent of the neutrino flavor $x$, counting all neutrinos above the 2.22 MeV
breakup threshold.  As the only detectable signal of the reaction is the produced neutron,
this channel placed very stringent constraints on the radioactive cleanliness
of the detector.  The third channel is the ES reaction of
conventional water detectors,
\begin{equation}
\nu_x + e^- \rightarrow \nu_x^\prime + e^{- \prime}.
\end{equation}

Operations were carried out in three phases.  SNO I operated with pure heavy water.
The NC-channel neutrons can capture on deuterium,
producing 6.25-MeV gamma rays that Compton scatter off electrons, producing
light for recoils above the Cherenkov threshold.
SNO I operations covered 306.4 live days from November 1999 through
May 2001.  Two analyses were performed based on the assumption of an 
undistorted $^8$B spectrum, using electron kinetic energy thresholds of 6.75
and 5 MeV, respectively.  The second analysis thus included NC events.

In SNO II two tons of purified NaCl were dissolved in the water, so that
$^{35}$Cl(n,$\gamma$) would become the dominant neutron sink.  This reaction
increases the capture rate and energy release.  Data were accumulated for 391.4
live days from July 2001 through August 2003.  Detector calibrations completed
in SNO I were repeated and extended in SNO II, including new checks involving
the introduction of beta-gamma sources that could 
lead to photo-disintegration of deuterium and the use of a
$^{252}$Cf neutron source to determine the neutron detection efficiency. The analysis
was performed for a kinetic energy threshold of 5.5 MeV and treated the first 254.2 live
days of data as blind.  In addition, the $^8$B spectrum shape was not assumed,
but rather extracted from the analysis, using 0.5 MeV bins from 5.5 to 13.5 MeV, plus
an additional bin for events between 13.5 and 20 MeV. 

In the first two phases of SNO the CC, ES, and NC rates were determined by a statistical
analysis that decomposed the common signal, the Cherenkov light, into the three 
contributing components.  The analysis exploited distinguishing
angular correlations with respect to the Sun and
energy differences in the CC-, ES-, and NC-associated light.  In SNO III the separation of
the NC and CC/ES signals was done by direct counting of NC neutrons.   
The salt introduced in SNO II was removed by
reverse osmosis and ion exchange, and a month of data was taken to confirm that the
detector had been restored to the operating conditions of SNO I.  Then an array of
the specially designed 
$^3$He- and CF$_4$-filled gas proportional counters was installed for neutron detection 
by ${}^3$He(n,p)$^3$H.  This
neutral-current detection (NCD) array 
consisted of 40 strings of proportional counters, ranging in length from 9 to 11 meters,
that were anchored to the inner surface of the acrylic vessel, forming a lattice on a 
one-meter grid. 

Between November 2004 and November 2006 385.17 live days of SNO III data were
taken.  Extensive calibrations of both the NCD and PMT arrays were made, utilizing
various neutron and gamma-ray sources, in order to calibrate the effectiveness of
the neutron detection and the impact of array installation on detector behavior.  The 
array was exploited to characterize neutron backgrounds within the detector,
including the distribution and isotopic composition of background sources.
During solar neutrino running, data were culled to eliminate
strings that exhibited mechanical or electrical faults, or runs (operational periods of
at least 30 minutes) when any array abnormalities were observed.  A blind analysis
of the remaining data was then performed.  The neutrino spectrum was again determined
from the CC and ES data, not assumed. 

The SNO I/II and SNO III results are in generally good agreement, and both separately and in combination established 
\begin{enumerate}
\item  A total flux of active neutrinos from ${}^8$B decay of $\phi_\mathrm{NC}(\nu_\mathrm{active})=(5.25 \pm 0.16 \mathrm{(stat)}^{+0.11}_{-0.13} \mathrm{(syst)}) \times 10^6 \mathrm{/cm^2/s}$, in good agreement with
SSM predictions, and $\phi_\mathrm{CC}(\nu_e) \sim$ 0.34 $\phi_\mathrm{NC}(\nu_\mathrm{active})$; and
\item The absence of statistically significant day-night effects or spectral distortions in
the portion of the $^8$B neutrino spectrum above $\sim$ 5 MeV;
\end{enumerate}

\subsection{Borexino}
The Borexino experiment \citep{BorexinoNIM}, located in the Gran Sasso Laboratory at an effective depth of about 3.0 
km.w.e.,
is the first to measure low-energy ($<$ 1 MeV) solar neutrino events in real time.  The
detector is housed within a 16.9m domed tank containing an outer layer of ultrapure water that provides
shielding against external neutrons and gamma rays.  At the inner edge of the water a stainless steel 
sphere serves as a support structure for an array of photomultiplier tubes that view both the inner
detector and the outer water shield, so that the Cherenkov light emitted by muons passing through
the water can be used to veto those events.  Within the steel sphere there are two regions, separated
by thin nylon vessels, containing high-purity buffer liquid, within which is sequestered a central volume
of 278 tons of organic scintillator.  The fiducial volume consists of $\sim$ 100 tons of the
liquid scintillator at the very center of the detector.  Scintillation light produced by recoil
electrons after ES events is the solar neutrino signal.  The 862 keV $^7$Be 
neutrinos produce a recoil electron spectrum with a distinctive cut-off edge 
at 665 keV.

The Borexino Collaboration reported results in 2008 and 2011 constraining the fluxes of  three low-energy
solar neutrino branches  \citep{Borexino,Borexinopep}:
\begin{enumerate}
\item A $^7$Be solar rate equivalent to an unoscillated flux of
$(3.10 \pm 0.15) \times 10^9$/cm$^2$s, or about 62\% of the GS98-SFII SSM central value;
\item A ES rate for $^8$B neutrinos, based on a integration above 3 MeV, corresponding to an
equivalent flux of $\phi^\mathrm{ES}(^8\mathrm{B}) = (2.4 \pm 0.4 \pm 0.1) \times 10^6/\mathrm{cm}^2\mathrm{s}$ \citep{Bellini10}, less precise
than but in good agreement with SNO and Super-Kamiokande results. [A similar result has been
obtained by the KamLAND collaboration,  $\phi^\mathrm{ES}(^8\mathrm{B}) = (2.77 \pm 0.26 \pm 0.32) \times 10^6/\mathrm{cm}^2\mathrm{s}$ from events above their 5.5 MeV analysis threshold \citep{KamLAND8}.]
\item The first direct, exclusive determination of the pep flux,
$(1.6 \pm 0.3) \times 10^8$/cm$^2$s (95\% c.l.); and
\item A limit on the CNO neutrino flux, $\phi_\mathrm{CNO} < 7.7 \times 10^8$/cm$^2$s
at 95\% c.l.
\end{enumerate}
The Borexino $^7$Be measurement places an important constraint on matter effects in neutrino oscillations, as
this line lies in a region dominated by vacuum oscillations, while the Super-Kamiokande and SNO
measurements are done in the matter-dominated region.

\section{NEW NEUTRINO PROPERTIES}
\label{sec:four}
The results just described have been addressed in global analyses that extract from the experiments
constraints on neutrino and solar properties.  Before describing such analyses, we discuss some of the 
associated weak interactions issues.
As previously noted, by 1994 Kamiokande II/III had finished operations, confirming the
neutrino deficit that Davis, Harmer, and Hoffman had first discovered 26 years earlier, and the SAGE
and GALLEX experiments had converged on a counting rate very close to the minimum astronomical
value of 78 SNU.  
The pattern of pp, $^7$Be, and $^8$B fluxes that emerged from analyses of the three early experiments (see Eq. (\ref{eq:fluxpattern})) was inconsistent with possible SSM variations altering $T_C$
(see figures by \cite{Castellani95} and \cite{HBL} included in \cite{Haxton95}) and improbable in
model-independent analyses that assumed only undistorted neutrino spectra \citep{HBL,HR}.   The agreement
between the SSM sound speed profile and that deduced from helioseismology also made it more
difficult to motivate SSM changes. 

A variety of new particle physics solutions to the solar neutrino problem had been suggested over the
years, including vacuum and matter-enhanced neutrino oscillations, neutrino decay \citep{BCY72}, and  weakly interacting massive particles 
(WIMPs) that might be bound in the Sun and consequently contribute to energy transport \citep{FG85,SP85}.
In addition to the standard MSW scenario, other oscillation effects in matter were explored, including
spin-flavor resonances driven by neutrino magnetic moments \citep{LM88,Akhmedov88},
parametric density fluctuations \citep{SK87,KS89}, contributions to the MSW potential from currents \citep{HZ91}, and
depolarization in the stochastic fluctuation limit \citep{LB94}.

Of these and other possibilities, the MSW mechanism drew the most attention because of its minimal requirements,
neutrino masses and a vacuum mixing angle $\theta_v \gtrsim 10^{-4}$.   While neutrinos are massless in the
SM, and consequently cannot oscillate, nonzero masses arise in most
extensions of the model.   Small weak interaction mixing angles were already familiar from the
analogous quark mixing matrix.

\subsection{Oscillation Basics: The Vacuum Case}
The current laboratory (tritium $\beta$ decay) limit on the $\bar{\nu}_e$ mass is 2.3 eV, though an effort
is underway to substantially improve this bound \citep{Katrin1,Katrin2}.  
Cosmological analyses variously limit  the sum over mass eigenstates to $\sum_i m_\nu(i) \lesssim 0.2-0.6$ eV \citep{Abazajian}.

Two types of neutrino mass terms can be added to the SM.
Neutrinos can have Dirac masses, 
analogous to those of other SM fermions, if the SM is enlarged to include a right-handed
neutrino field.  Because 
neutrinos lack charges or other additively conserved quantum numbers, 
lepton-number-violating Majorana mass terms can also be added,
$\overline{ \nu_L^c} m_L \nu_L$ and $\overline{\nu_R^c} m_R \nu_R$, where the
former is the only dimension-five operator that can be constructed in the SM.  
(The subscripts $L$ and $R$ denote left- and right-hand projections of the neutrino field $\nu$, and
the superscript $c$ denotes charge conjugation.) 

In the seesaw mechanism \citep{Seesaw1,Seesaw2,Seesaw3} the Dirac and Majorana mass terms are combined in a manner that provides
an attractive explanation for light neutrinos,
\[ M_\nu \sim \left( \begin{array}{cc} 0 & m_D \\ m_D^T & m_R \end{array} \right), \]
where $m_L \sim 0$ in part because of double beta decay constraints.  When diagonalized,
the matrix yields heavy and light neutrino mass eigenstates,
\[ m_H \sim m_R~~~~~~~~~~~~~~~m_L \sim m_D  {m_D \over m_R} \]
with the latter related to the typical Dirac mass of the SM by the coefficient $m_D/m_R$.   If we assume 
the scale of the new physics which $m_R$ represents is $\gg m_D$, then a candidate small parameter
$m_D/m_R$ is available to explain why neutrinos are so much lighter than other SM fermions.
Small neutrino masses are thus explained as a consequence of the
scale $m_R$ of new physics beyond the SM.

Neutrinos of definite mass are the stationary
states for free propagation, while neutrino flavor eigenstates are produced in weak interactions.  
Simplifying here to two flavors, the relationship of the flavor $\{\nu_e,\nu_\mu\}$ and mass $\{\nu_1,\nu_2\}$ eigenstates 
can be described by a single vacuum mixing angle $\theta_v$,
\begin{equation}
\nu_e = \cos{\theta_v} |\nu_1 \rangle + \sin{\theta_v} | \nu_2 \rangle ~~~~~~~~\nu_\mu = -\sin{\theta_v} |\nu_1 \rangle + \cos{\theta_v} | \nu_2 \rangle.
\end{equation}
Consequently, an arbitrary
initial state $|\nu(t=0) \rangle = a_e(t=0) |\nu_e \rangle + a_\mu(t=0) |\nu_\mu \rangle$ of momentum $k \sim E$, as it propagates
downstream, evolves according to
\begin{equation}
i  {d \over dt} \left( \begin{array}{c} a_e \\ a_\mu \end{array} \right) = { 1 \over 4 E} \left( \begin{array}{cc}
-\delta m_{21}^2 \cos{2 \theta_v} & \delta m_{21}^2 \sin{2 \theta_v} \\ \delta m_{21}^2 \sin{2 \theta_v} & \delta m_{21}^2 \cos{2 \theta_v} 
\end{array} \right)  \left( \begin{array}{c} a_e \\ a_\mu \end{array} \right).
\label{eq:vacuum}
\end{equation}
where an average overall wave function phase has been removed from the neutrino mass matrix (represented
in the flavor basis).  For the special case of a $\nu_e$ produced at time $t=0$, the solution of this
equation yields
\begin{equation}
P_{\nu_e}(t) = | \langle \nu_e | \nu(t) \rangle|^2 = 1 -\sin^2{2 \theta_v} \sin^2{\left( { \delta m_{21}^2 t \over 4 E} \right)} \rightarrow 1-{1 \over 2} \sin^2{2 \theta_v}
\end{equation}
where the downstream oscillation depends on the difference $\delta m_{21}^2 \equiv m_2^2-m_1^2$.  (If this
problem is done properly with wave packets, the oscillation persists until the two mass components separate
spatially, yielding the asymptotic result on the right.)   The
oscillation length $ L_0 = 4 \pi \hbar c E / \delta m_{21}^2 c^4$
is shorter than the Earth-Sun distance for a typical solar neutrino of energy $\sim$ 1 MeV provided
$\delta m_{21}^2 \gtrsim 1.6 \times 10^{-11}$ eV$^2$.  Thus solar neutrinos are
interesting for oscillation studies because of their sensitivity to extremely small neutrino
mass differences.

\subsection{Oscillation Basics: Matter and the MSW Mechanism}
  
\cite{MS1985,MS1986} showed that the
density dependence of the neutrino effective mass, a phenomenon
first discussed by \cite{Wolfa,Wolfb}, could greatly enhance
oscillation probabilities.  Their original numerical work was
soon understood analytically as a 
consequence of level crossing:  a neutrino produced in the core as a
$\nu_e$ is adiabatically transformed
into a $\nu_\mu$ by traversing a critical solar density 
where the $\nu_e$ and $\nu_\mu$ effective masses cross.
It became clear that the Sun is not only an excellent 
neutrino source, but also a natural regenerator for
enhancing the effects of flavor mixing.

Equation (\ref{eq:vacuum}) describing vacuum oscillations is
altered in matter
\begin{equation}
i {d \over dx} \left( \matrix { a_{\textstyle e} \cr
a_{\textstyle \mu} \cr} \right) = {1 \over 4E} \left ( \matrix{
2E \sqrt2 G_F \rho(x) - \delta m_{21}^2 \cos 2 \theta_{\textstyle v}
~~~~~~\delta m_{21}^2\sin
2\theta_{\textstyle v} \cr 
\delta m_{21}^2\sin 2 \theta_{\textstyle v} ~~~ -2E \sqrt2 G_F \rho(x) +
\delta m_{21}^2
\cos 2\theta_{\textstyle v} \cr} \right) \left( \matrix {
a_{\textstyle e} \cr
a_{\textstyle \mu} \cr} \right) 
\label{eq:matter}
\end{equation}
where G$_F$ is the weak coupling constant and $\rho (x)$ the solar
electron number density. 
The new contribution to the difference in diagonal elements, $4 E \sqrt2 G_F \rho(x)$, 
represents the effective contribution to $m^2_\nu$  that arises 
from neutrino-electron scattering.  The indices of refraction
of electron and muon neutrinos differ because the former
scatter via charged and neutral currents, while the latter 
have only neutral current interactions.  
For $\theta_v \lesssim \pi/4$ -- the ``normal hierarchy" where the lighter mass eigenstate 
makes the larger contribution to $\nu_e$ in vacuum --
the matter and vacuum contributions to the diagonal elements of Eq. (\ref{eq:matter})
have opposite signs.

 We can diagonalize the right-hand side of Eq. (\ref{eq:matter}) to
 determine the heavy and light local mass eigenstates and eigenvalues $m_H(x)$ and $m_L(x)$,
 functions of $\rho(x)$
\begin{equation}
|\nu_L (x)\rangle = \cos \theta (x)|\nu_e\rangle - \sin \theta 
(x)|\nu_\mu\rangle ~~~~~~
|\nu_H(x)\rangle = \sin \theta (x)|\nu_e\rangle + \cos \theta (x)|\nu_\mu 
\rangle 
\end{equation}
where the local mixing angle
\begin{equation}
\sin 2 \theta (x)  = {\sin 2 \theta_{\textstyle v} \over \sqrt{X^2 (x) + \sin^2
2\theta_{\textstyle v}}}~~~~~
\cos 2\theta (x)  = {-X (x) \over \sqrt{X^2 (x) + \sin^2 2\theta_{\textstyle
v}}} 
\label{eq:angles}
\end{equation}
depends on $X(x) = 2 \sqrt2G_F \rho(x) E/\delta m_{21}^2 - \cos 2\theta_{\textstyle v}$.
Unlike the vacuum case, these are stationary states for propagation only 
if $\rho(x)$ is constant .  Otherwise, defining
$|\nu (x) \rangle = a_H(x)|\nu_H(x)\rangle + a_L(x)|\nu_L(x)\rangle$,
Eq. (\ref{eq:matter}) becomes
\begin{equation}
i {d \over dx} \pmatrix{
a_H \cr
a_L \cr} = {1 \over 4 E} \pmatrix {
\lambda(x) & i \alpha (x) \cr
-i \alpha (x) & - \lambda (x) \cr }
\pmatrix
{a_H \cr
a_L }.
\label{eq:localmass}
\end{equation}
The splitting of the local mass eigenstates and the local oscillation length are
\begin{equation}
\lambda (x) = {\delta m_{21}^2} \sqrt{X^2 (x) + \sin^2 2\theta_{\textstyle v}} ~~~~~~~~L_0(x) = {4 \pi \hbar c E \over \lambda(x) c^4}
\end{equation}
while eigenstate mixing is governed by the density gradient
\begin{equation}
\alpha (x) = \left({4E^2 \over \delta m_{21}^2}\right)
 \, {\sqrt2 \, G_F {d \over dx}
\rho(x)
\sin 2 \theta_{\textstyle v} \over X^2 (x) + \sin^2 2 \theta_{\textstyle v}}.
\end{equation}
The splitting achieves
its minimum value, ${2 \delta m_{21}^2} \sin 2 \theta_v$, at a critical density $\rho_c =
\rho (x_c)$ where $X(x) \rightarrow 0$,
\begin{equation}
2 \sqrt2 E G_F \rho_c = \delta m_{21}^2 \cos 2 \theta_v.
\label{eq:critical}
\end{equation}
The diagonal elements of the original flavor matrix 
of Eq. (\ref{eq:matter}) cross at $\rho_c$.  

\begin{figure}[htb]
\includegraphics[width=13.4cm]{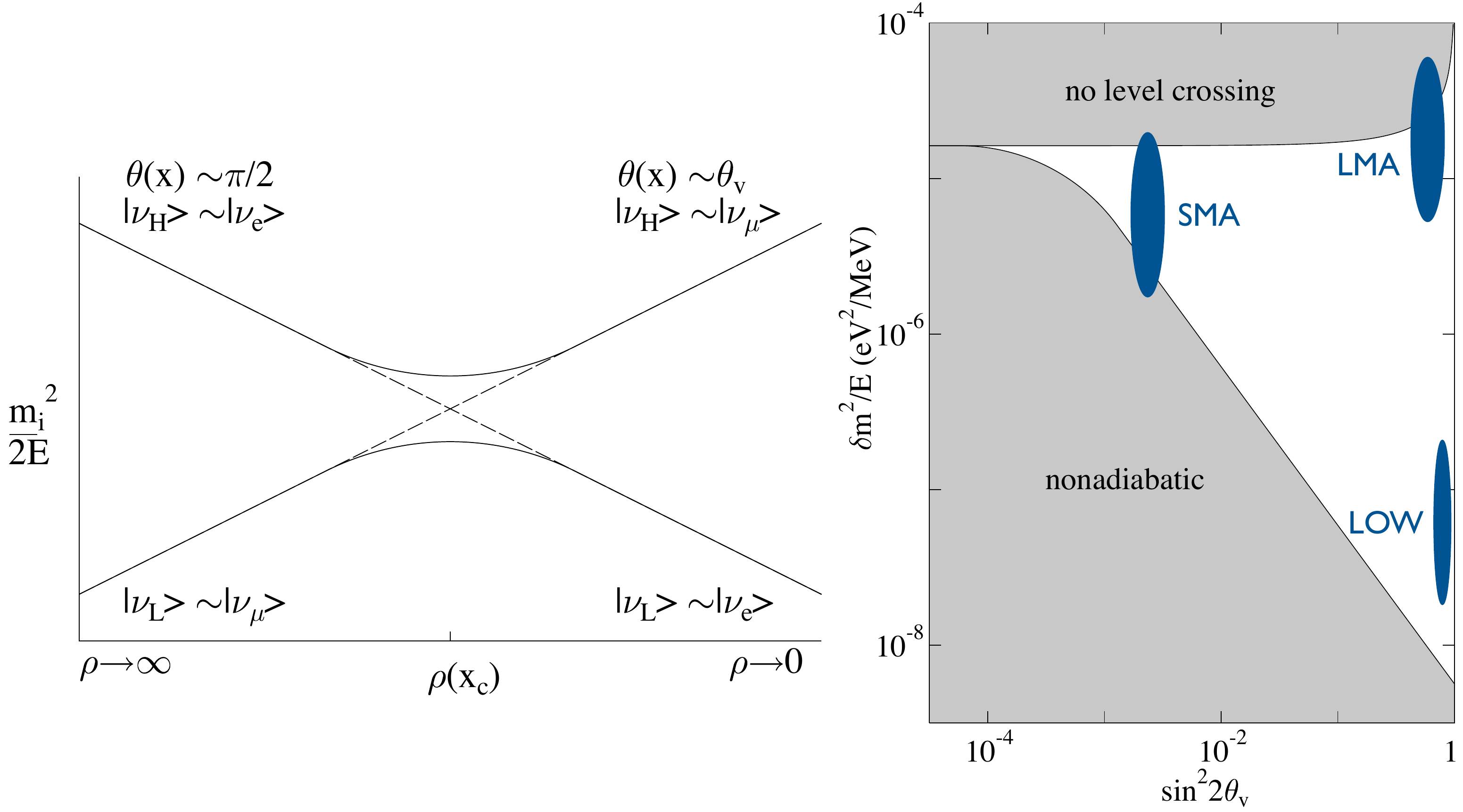}
\caption{(Color online) Left: A schematic illustration of the MSW crossing for a normal
hierarchy and small $\theta_v$.  The dashed 
lines -- the electron-electron and muon-muon diagonal
elements of the $m_\nu^2$ matrix --
intersect at the level-crossing density $\rho_c$.
The solid lines are the trajectories of the light and heavy
local mass eigenstates.  A $\nu_e$ produced at high density as $\sim \nu_H$ will, under 
adiabatic propagation, remain $\sim \nu_H$, exiting the Sun as $\sim \nu_\mu$. 
Right: the white ``MSW triangle" is the region where a level crossing occurs and propagation is adiabatic,
producing strong $\nu_e \rightarrow \nu_\mu$ conversion.  Regions of three possible MSW solutions 
frequently discussed in the 1990s -- SMA, LMA, and LOW -- are depicted by the blue ellipses.}
\label{fig:MSW}
\end{figure}

The crux of the MSW mechanism is the adiabatic crossing
of the critical density, illustrated in the left panel of Fig. \ref{fig:MSW}.
The adiabatic condition is determined by the requirement
$\gamma (x) = \left|{\lambda (x) \over \alpha (x)}\right|  \gg 1$, which allows one to treat  Eq. (\ref{eq:localmass}) as
diagonal.  This condition becomes particularly stringent near the crossing point,
\begin{equation}
\gamma_c \equiv \gamma (x_c) = {\sin^2 2\theta_v \over \cos 2\theta_v} \, {\delta
m_{21}^2 \over 2 E} \, {1 \over \left|{1 \over \rho_c} {d \rho (x) \over dx}|_{x =
x_c}\right|}  = 2 \pi \tan 2\theta_v~ {H_c \over L_c}
\gg 1,
\end{equation}
where $H_c$ and $L_c$ are the solar density scale height and local oscillation
length at $x_c$.  If the $H_c \gg L_c$,
Eq. (\ref{eq:localmass}) then yields \citep{Bethe86}
\begin{equation}
P^{\rm adiab}_{\nu_e} = {1 \over 2} + {1 \over 2} \cos 2 \theta_v \cos 2
\theta_i 
\label{eq:adiabatic}
\end{equation}
where $\theta_i = \theta (x_i)$ is the local mixing angle at the density where
the neutrino is produced.  The adiabatic solution depends on the local mixing
angles where the neutrino begins ($\theta_i$, solar core) and ends ($\theta_v$, in
vacuum) its propagation.

For illustration, consider the case of a small $\theta_v \sim 0$.   A solar $\nu_e$
created in the high-density solar core is then nearly identical to the heavy-mass eigenstate ($\theta_i \sim \pi/2$),
provided the vacuum mass difference between the eigenstates is not too large
(see  Eq. (\ref{eq:angles}).  If the subsequent propagation 
is adiabatic,
the neutrino remains on the heavy-mass trajectory, crossing the critical density ($\theta(x_c) = \pi/4$),
and finally exiting the Sun.  But in vacuum the heavy mass
eigenstate is $\sim \nu_\mu$: a nearly complete flavor change, $\nu_e \rightarrow \nu_\mu$, has occurred,
through an adiabatic rotation of the local oscillation angle
from $\theta_i \sim \pi/2$ to $\theta_f=\theta_v$ during propagation.

If the adiabatic condition is not satisfied, e.g., $\gamma_c \lesssim 1$, 
an accurate analytic solution can still be obtained \citep{Haxton86,Parke86}.  As we have seen, the 
nonadiabatic behavior is governed by the density
scale height at $x_c$.  One can replace the actual solar density by an effective one, e.g.,
a linear density ``wedge" that has the correct derivative at $x_c$ (thereby incorporating the effects of the 
density gradient at  the most sensitive point), while also starting and ending at the appropriate initial and final densities (thereby 
also building in the adiabatic limit).
The resulting generalization of Eq. (\ref{eq:adiabatic}) is
\begin{equation}
P_{\nu_e} = {1 \over 2} + {1 \over 2} \cos 2 \theta_v \cos 2 \theta_i ( 1 - 
2P_{\rm {hop}}) ~~~~~~~P_{\rm {hop}} \equiv e^{- \pi \gamma_c/2}
\label{eq:hop}
\end{equation}
where P$_{\rm {hop}}$, the Landau-Zener probability of hopping from the heavy mass
trajectory to the light trajectory on traversing $x_c$, vanishes in the highly adiabatic limit, $\gamma_c \gg 1$
(so that Eq. (\ref{eq:hop}) reduces to Eq. (\ref{eq:adiabatic})).
When the crossing becomes highly nonadiabatic ($\gamma_c \ll 1$ ),
P$_{\rm {hop}} \rightarrow 1$: the neutrino
exits the Sun on the light mass trajectory, which for small mixing angles
means it remains $\sim \nu_e$.

Thus strong           
conversion of solar neutrinos is expected when 1) the propagation is adiabatic ($\gamma_c \gtrsim 1$) and 2)  
there is a level 
crossing (there is enough matter at the $\nu_e$ production point that $\nu_e(x_i) \sim \nu_H(x_i)$).
The right panel of Fig. \ref{fig:MSW} shows the  white triangle of
parameters in the $\delta m_{21}^2/ E - \sin^2 2\theta_v$ plane where 
both constraints are satisfied.  Within this triangle, strong conversion occurs.
One can envision superimposing on this triangle the spectrum of solar neutrinos, plotted as a 
function of $\delta m_{21}^2/E$ for some choice of $\delta m_{21}^2$ and $\theta_v$.  
Depending on how that spectrum is positioned vertically (a function of $\delta m_{21}^2$)
or horizontally (a function of $\theta_v$) one can alter the resulting spectrum
of $\nu_e$s in several characteristic ways, for example, suppressing the low-energy
or high-energy neutrinos preferentially, or even (in the case of small mixing angles)
those of intermediate energy.  

In early fits to the neutrino data three potential MSW solutions were frequently discussed,
designated by SMA (small mixing angle: $\delta m_{21}^2 \sim 5.4 \times 10^{-6} \mathrm{~eV},~
\sin^2{2 \theta} \sim 0.006$), LMA (large mixing angle: $\delta m_{21}^2 \sim 1.8 \times 10^{-5} \mathrm{~eV},~
\sin^2{2 \theta} \sim 0.76$), and LOW (low probability, low mass:
$\delta m_{21}^2 \sim 7.9 \times 10^{-8} \mathrm{~eV},~
\sin^2{2 \theta} \sim 0.96$).  (The parameter values are taken from
\cite{BKS} and are representative of fits done at that time.)  These solutions are indicated schematically by the colored regions
in  Fig.  \ref{fig:MSW}.  The 
solution consistent with the solar neutrino data proved, ironically, to be the LMA solution --
not the SMA solution where matter effects so greatly enhance the oscillations.

\section{GLOBAL ANALYSES AND NEUTRINO PROPERTIES}
\label{sec:five}
Neutrino physics has made great progress in the last 15 years, as reactor and accelerator neutrino
experiments have added new information to that obtained from solar and atmospheric
neutrino experiments.  The three mixing angles of the 3$\times$3 neutrino
mass matrix, the magnitudes of the two mass differences, and from solar neutrino 
experiments the sign of one of these have all been determined.  The phenomena that can be
explored with solar neutrinos were illustrated previously for the
two-flavor case:  flavor oscillations, affected by matter, not only alter fluxes, but
lead to distinctive spectral distortions, and may produce day-night differences due to neutrino passage through
the Earth.   The various experimental collaborations as well as independent groups have
developed global analysis methods to analyze solar neutrino experiments, taking into
account the constraints other recent measurements have imposed.  Here we summarize the
conclusions of such analyses, relying particularly on work done by the Bari and Valencia groups.

\subsection{Vacuum Mixing Angles and Mass$^2$ Differences}
In the SM case of three light neutrino flavors, the relationship between flavor \{$\nu_e,\nu_\mu,\nu_\tau$ \} and mass \{ $\nu_1,\nu_2,\nu_3$ \} eigenstates
is described by the  PMNS matrix \citep{MNS,Pontecorvo67}
\begin{equation}
\left( \begin{array}{c} | \nu_e \rangle \\ | \nu_\mu \rangle \\ | \nu_\tau \rangle \end{array} \right) =
\left( \begin{array}{ccc} c_{12} c_{13} & s_{12} c_{13} & s_{13} e^{-i \delta} \\
-s_{12}c_{23}-c_{12}s_{23} s_{13} e^{i \delta} & c_{12}c_{23}-s_{12}s_{23}s_{13} e^{i \delta} & s_{23} c_{13} \\
s_{12}s_{23}-c_{12}c_{23}s_{13} e^{i \delta} & -c_{12}s_{23}-s_{12}c_{23}s_{13}e^{i \delta} & c_{23} c_{13} \end{array}
\right) \left( \begin{array}{c} e^{i \alpha_1 /2} | \nu_1\rangle \\ e^{i \alpha_2 /2} | \nu_2 \rangle \\ | \nu_3 \rangle \end{array} \right)
\end{equation}
where $c_{ij} \equiv \cos{\theta_{ij}}$ and $s_{ij} \equiv \sin{\theta_{ij}}$.  This matrix depends on
three mixing angles $\theta_{12}$, $\theta_{13}$, and $\theta_{23}$, of which the first and last are the
dominant angles for solar and atmospheric oscillations, respectively; a Dirac phase $\delta$ that
can induce CP-violating differences in the oscillation probabilities for
conjugate channels such as $\nu_\mu \rightarrow \nu_e$ versus
$\bar{\nu}_\mu \rightarrow \bar{\nu}_e$; and two Majorana phases $\alpha_1$ and $\alpha_2$
that will affect the interference among mass eigenstates in the effective neutrino mass probed
in the lepton-number-violating process of neutrinoless double $\beta$ decay.

\begin{table}
\caption{Results from global $3\nu$ analyses including data through Neutrino2012.  }
\vspace{0.4cm}
\scalebox{0.8}{
\begin{tabular}{|l|c|c|c|c|c|c|}
\hline \hline
 & \multicolumn{3}{c|} {Bari Analysis \citep{Bari}} & \multicolumn{3}{c|} {Valencia Analysis \citep{Valencia}} \\
\hline
 Parameter/hierarchy &Best $1\sigma$ Fit& $2\sigma$ Range &  $3\sigma$ Range &Best $1\sigma$ Fit& $2\sigma$ Range & $3\sigma$ Range  \\
 \hline
$\delta m_{21}^2 (10^{-5} \mathrm{eV}^2)$ &  7.54$^{+0.26}_{-0.22}$ &7.15 $\leftrightarrow$ 8.00&6.99 $\leftrightarrow$ 8.18&7.62$\pm$0.19&7.27 $\leftrightarrow$ 8.01&7.12 $\leftrightarrow$ 8.20\\
$\delta m_{31}^2 (10^{-3} \mathrm{eV}^2)$ NH&
$2.47^{+0.06}_{-0.10}$&2.31 $\leftrightarrow$ 2.59& 2.23 $\leftrightarrow$ 2.66&$2.55^{+0.06}_{-0.09}$&2.38 $\leftrightarrow$ 2.68&2.31 $\leftrightarrow$ 2.74\\
\hspace{2.6cm} IH&$-(2.38^{+0.07}_{-0.11})$&$-$(2.22 $\leftrightarrow$ 2.49)&$-$(2.13 $\leftrightarrow$ 2.57)& $-(2.43^{+0.07}_{-0.06})$&$-$(2.29 $\leftrightarrow$ 2.58)&$-$(2.21 $\leftrightarrow$ 2.64)
 \\
$\sin^2{\theta_{12}}$ & $0.307^{+0.018}_{-0.016}$ & 0.275 $\leftrightarrow$ 0.342 & 0.259 $\leftrightarrow$ 0.359 &
$0.320^{+0.016}_{-0.017}$ & 0.29 $\leftrightarrow$ 0.35 & 0.27 $\leftrightarrow$ 0.37 \\
$\sin^2{\theta_{23}}$ \hspace{1.2cm} NH & $0.386^{+0.024}_{-0.021}$ & 0.348 $\leftrightarrow$ 0.448 & 0.331 $\leftrightarrow$ 0.637 &$ \left\{ \begin{array}{c} 0.613^{+0.022}_{-0.040} \\  0.427^{+0.034}_{-0.027} \end{array} \right.$ & 0.38 $\leftrightarrow$ 0.66 & 0.36 $\leftrightarrow$ 0.68 \\
\hspace{2.6cm} IH & $0.392^{+0.039}_{-0.022}$ & $ \left\{ \begin{array}{c} 0.353 \leftrightarrow 0.484 \\ 0.543 \leftrightarrow 0.641 \end{array} \right.$ & 0.335 $\leftrightarrow$ 0.663 &
$0.600^{+0.026}_{-0.031}$ & 0.39 $\leftrightarrow$ 0.65 & 0.37 $\leftrightarrow$ 0.67 \\
$\sin^2{\theta_{13}}$ \hspace{1.2cm} NH & $0.0241\pm 0.0025$ & 0.0193 $\leftrightarrow$ 0.0290 & 0.0169 $\leftrightarrow$ 0.0313 &
$0.0246^{+0.0029}_{-0.0028}$ & 0.019 $\leftrightarrow$ 0.030 & 0.017 $\leftrightarrow$ 0.033 \\
\hspace{2.6cm} IH & $0.0244^{+0.0023}_{-0.0025}$ & 0.0194 $\leftrightarrow$ 0.0291 & 0.0171 $\leftrightarrow$ 0.0315 &
$0.0250^{+0.0026}_{-0.0027}$ & 0.020 $\leftrightarrow$ 0.030 & 0.017 $\leftrightarrow$ 0.033 \\
\hline \hline																					
\end{tabular}
\label{tab:global}	
}
\end{table}

It became apparent in early analyses that combined solar and reactor neutrino data in two-flavor
analyses that there was some hint of the third flavor, a nonzero $\theta_{13}$.  The KamLAND Collaboration
analysis employed the 
three-flavor $\nu_e$ survival probability of \cite{Fogli00} in which the
influence of $\theta_{13}$ in modifying the two-flavor result is explicit,
\begin{equation} 
P_{ee}^{3 \nu} = \cos^4{\theta_{13}} ~\tilde{P}_{ee}^{2 \nu} + \sin^4{\theta_{13}}
\end{equation}
where $\tilde{P}_{ee}^{2 \nu}$ is the two-flavor survival probability in matter evaluated for the
modified electron density $\rho(x) \rightarrow \rho(x) \cos^2{\theta_{13}}$.   The analysis
yielded $\sin^2{\theta_{13}} = 0.020 \pm 0.016$  \citep{KamLAND11}, a result consistent with
the long-baseline $\nu_e$ appearance results announced shortly afterwards,
$ 0.008 \lesssim \sin^2{\theta_{13}} \lesssim 0.094$ \citep{T2K} and  $0.003 \lesssim \sin^2{\theta_{13}} \lesssim 0.038$
\citep{MINOS}.  In 2012 results from reactor $\bar{\nu}_e$ disappearance experiments 
became available, yielding $\sin^2{\theta_{13}} = 0.022 \pm 0.011 (\mathrm{stat}) \pm 0.008 (\mathrm{sys})$ \citep{DC},
$\sin^2{\theta_{13}} = 0.0236 \pm 0.0042 (\mathrm{stat}) \pm 0.0013 (\mathrm{sys})$ \citep{DayaBay}, and
$\sin^2{\theta_{13}} = 0.0291 \pm 0.0035 (\mathrm{stat}) \pm 0.0051 (\mathrm{sys})$ \citep{Reno}.  The latter
two results, because of their precision, effectively remove a degree of freedom from three-flavor
solar neutrino analyses.

The mass differences and mixing angles from the global analyses of the Bari \citep{Bari} and Valencia \citep{Valencia}
groups, including experimental results through the Neutrino 2012 Conference, are shown in Table \ref{tab:global}.
The two analyses are generally in quite good agreement and yield (in degrees)
\begin{equation}
\theta_{12} \sim \left\{ \begin{array}{c} 33.6^{+1.1}_{-1.0}  \\ 34.4^{+1.0}_{-1.1}  \end{array} \right. 
~~~~~\theta_{13} \sim \left\{ \begin{array}{cl} 8.96^{+0.45}_{-0.51} & \mathrm{~~~~~~~~~~Bari} \\ 9.06^{+0.50}_{-0.57}
& \mathrm{~~~~~~~Valencia}\end{array}  \right.
\end{equation}
The agreement in the solar neutrino mass difference $\delta m_{21}$ is also excellent,
\begin{equation}
\delta m^2_{21} \sim \left\{ \begin{array}{cl} (7.54^{+0.26}_{-0.22}) \times 10^{-5}~\mathrm{eV}^2 & \mathrm{~~~~~~~~~~Bari} \\ (7.62^{+0.19}_{-0.19}) \times 10^{-5}~\mathrm{eV}^2
& \mathrm{~~~~~~~Valencia}\end{array}  \right.
\end{equation}

The values for $\theta_{12}$ and $\delta m_{21}^2$ 
lie in the LMA region of Fig. \ref{fig:MSW}.  $\delta m_{21}^2$ corresponds, for 10 MeV neutrinos, to an MSW crossing
density of $\sim$ 20 g/cm$^3$, or equivalently a solar radius of 
$r \sim 0.24 R_\odot$, the outer edge of the Sun's energy-producing core.
The crossing density for the atmospheric $\delta m_{31}^2$, again for 10 MeV neutrinos, is
$\sim 1.6 \times10^3$ g/cm$^3$.  Thus this crossing requires electron densities
far beyond those available in the Sun -- though typical of the carbon zone in the mantle
of a Type II supernova, where this second crossing plays a significant role.

\begin{figure}
\includegraphics[width=14cm]{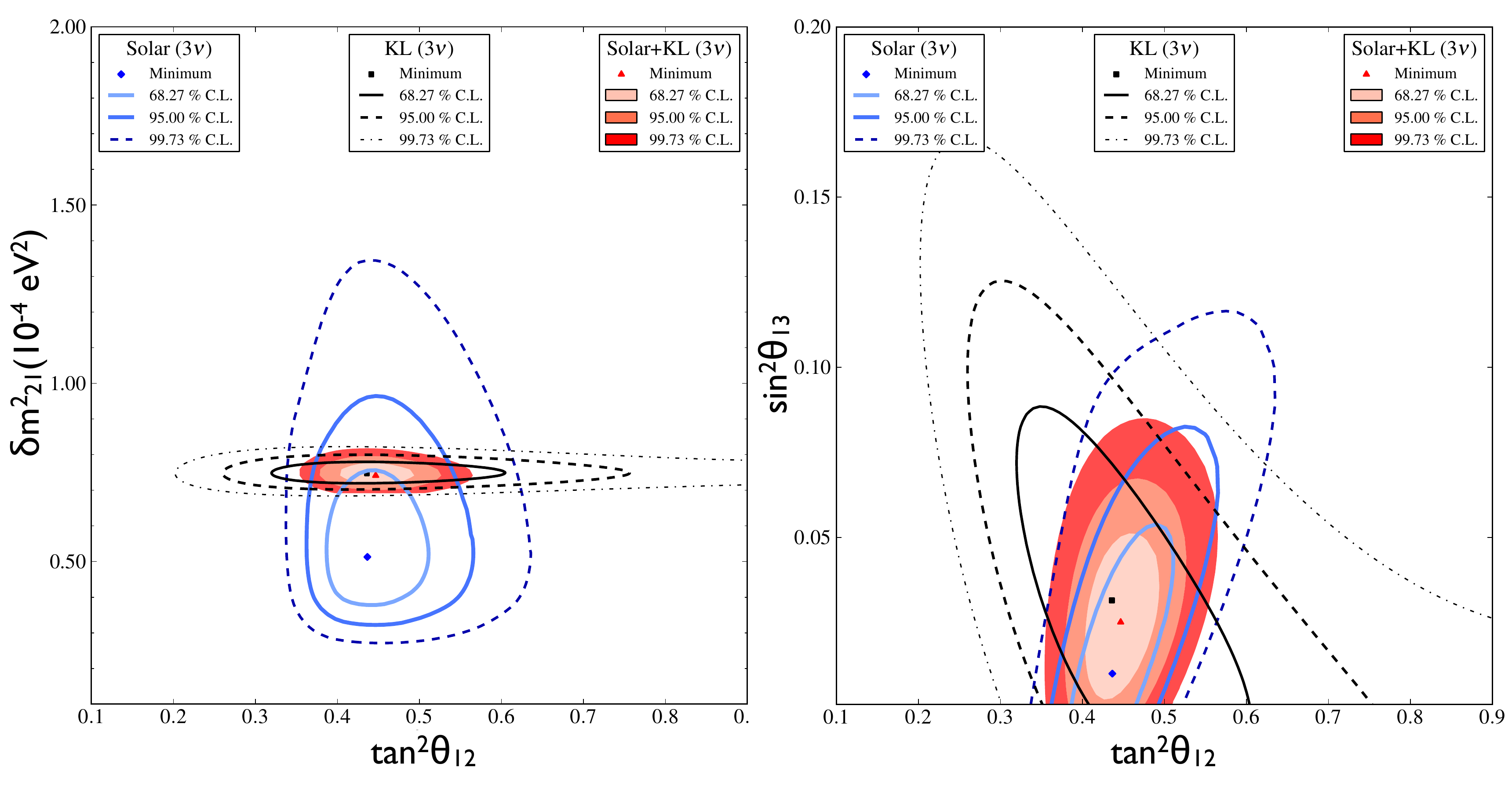}
\caption{(Color online) The three-flavor neutrino oscillation contours resulting from the SNO combined analysis:
the analysis employs only solar neutrino and KamLAND data, but the results are in excellent agreement with
the conclusions from global analyses that include recent reactor- and accelerator-neutrino constraints on $\theta_{13}$.
 From \cite{SNOCombined}.}
\label{fig:SNO3flavor}
\end{figure}

These global analysis results can be compared with those from the recent SNO three-flavor combined analysis, which
used all available solar neutrino data and the results from KamLAND.  This
analysis, summarized in Fig. \ref{fig:SNO3flavor}, gives at
$1\sigma$
\begin{equation}
\sin^2{\theta_{12}} = 0.308 \pm 0.014~~~\delta m_{21}^2 = (7.41^{+0.21}_{-0.19}) \times 10^{-5} \mathrm{~eV^2}~~~
\sin^2{\theta_{13}}=0.025^{+0.018}_{-0.015}
\end{equation}
These values are in excellent agreement with the corresponding $1\sigma$ Bari and Valencia results
of Table \ref{tab:global}: the SNO combined analysis and Bari best values match
particularly well.  The main consequence of the inclusion of new reactor and
accelerator results in the global analyses is a substantial reduction in the uncertainty on $\theta_{13}$.

\subsection{Spectral Distortions: LET Analyses and Borexino}
Characteristic spectral distortions are one of the signatures of oscillations in matter.  Rather fortuitously,
if one evaluates Eq. (\ref{eq:critical}) for the neutrino energy where the MSW critical density corresponds 
to the electron density at  the center of the Sun, $\rho \sim 6 \times 10^{25}/\mathrm{cm}^3$, one finds
$E^\nu_\mathrm{crit} \sim 1.9$ MeV, an energy in the center of the solar neutrino spectrum.  
Neutrinos below this energy
will not experience a level crossing on exiting the Sun, and thus will oscillate approximately as they
would in vacuum -- an average (two-flavor) survival probability of
\begin{equation}
P_{\nu_e}^\mathrm{vacuum} \sim 1 - {1 \over 2} \sin^2{2 \theta_{12}} \sim 0.57,
\end{equation}
using $\theta_{12} \sim 34^\circ$.  This can be compared to the matter-dominated survival
probability, appropriate for neutrinos much above the critical energy, 
\begin{equation}
P_{\nu_e}^\mathrm{high~density} \rightarrow \sin^2{\theta_{12}}  \sim 0.31.
\end{equation}
Most of the $^8$B neutrinos studied by SNO and Super-Kamiokande
undergo matter-enhanced oscillations.  The matter/vacuum transition predicted by the MSW
mechanism can be verified by comparing the survival probabilities of low-energy
(pp or $^7$Be) and high-energy ($^8$B) neutrinos.  
Alternatively, if the thresholds in SNO and Super-Kamiokande are
lowered sufficiently, spectral distortions will be detectable in the $^8$B 
spectrum:  low-energy $^8$B neutrinos coming from
the outer core will not experience a crossing, and thus will have a higher survival rate.

The flux of low-energy pp neutrinos is well constrained in global analyses because these
neutrinos dominate the SAGE and GALLEX/GNO counting rates.   (The need for an elevated survival probability for
these neutrinos was an important factor in early model-independent analyses that concluded
undistorted neutrino fluxes could not account for the data.)   Furthermore
Borexino has now provided a direct, exclusive measurement at a precise energy, 
corresponding to the 860 keV neutrinos from $^7$Be electron capture.

To probe lower energy $^8$B neutrinos, a joint re-analysis of Phase I and Phase II data from 
the Sudbury Neutrino Observatory was carried out with an effective kinetic energy threshold of
$T_\mathrm{eff} = 3.5$ MeV \citep{SNOLET}.  While
the low-energy threshold analysis (LETA)
had several motivations (e.g., the enlarged data set improved the overall
precision of the flux determinations), a principal goal was enhancing prospects for detection of
the predicted upturn in $P_{\nu_e}$ with decreasing neutrino energy.
An effort similar to the LETA analysis is now underway in Super-Kamiokande IV: preliminary results were
recently described by \cite{Smy}.

\begin{figure}
\includegraphics[width=13cm]{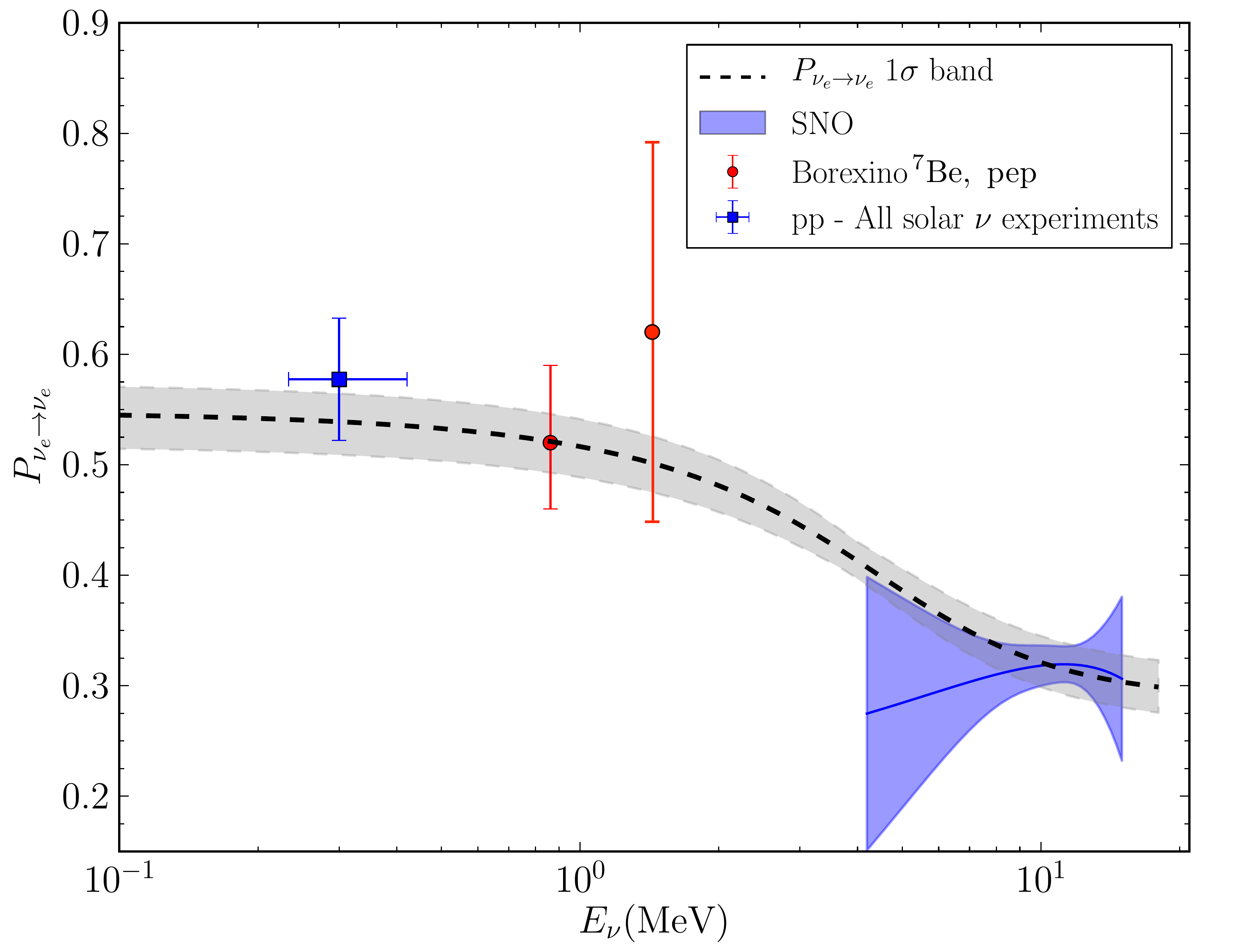}
\caption{(Color online) Survival probabilities $P_{\nu_e}$ for pp, pep, $^7$Be, and $^8$B neutrinos 
deduced from global solar neutrinos analyses, Borexino, and the SNO combined analysis, compared to
the MSW prediction, taking into account present uncertainties on mixing angles.  From
\cite{SNOCombined}, with pep result from \cite{Borexinopep} added.}
\label{fig:distortion}
\end{figure}

Figure \ref{fig:distortion} summarizes the data.  The pattern defined by the pp $\nu_e$ flux deduced
from global analyses, the $^7$Be $\nu_e$ flux derived from the Borexino ES measurement,
and the SNO results are generally in good agreement with the expected MSW survival probability.
However, while the SNO LETA band is 
compatible with the MSW prediction, the band's centroid trends away from the theory with
decreasing energy. 

\subsection{Day-Night Differences}
Two sources of time variation in neutrino rates are the annual $\sim 7\%$ modulation associated
with the 1.7\% eccentricity in the Earth's orbit around the Sun, and the daily variation associated
with terrestrial matter effects, which influence the night-time flux of up-going neutrinos.
Both effects have been the subject of careful experimental studies: see, e.g., \cite{SKIA,SNOPeriodicity}.
The integrated day-night asymmetry in neutrino detection rates
\begin{equation}
A_\mathrm{DN} \equiv {R_\mathrm{D} - R_\mathrm{N} \over {1 \over 2} (R_\mathrm{D} + R_\mathrm{N})},
\end{equation}
where $R_\mathrm{D}$ and $R_\mathrm{N}$ denote the day and night rates, is the quantity
most often studied to assess matter effects associated with solar neutrino passage through the Earth.  
In principle, similar differential quantities could be defined as functions of the neutrino energy and
zenith angle.  However, the detection of even the integrated difference $A_\mathrm{DN}$ is statistically challenging,
as the effect is expected to be only a few percent. 

$A_\mathrm{DN}$ provides an ``on-off" test where the matter effects 
can be measured directly, unlike the solar case where matter effects must be deduced from phenomena
such as spectral distortions.
The magnitude of the neutrino regeneration associated with passage through the Earth depends
on the neutrino energy, the assumed oscillation parameters, and, to some extent, 
detector location, as that determines the possible trajectories through the Earth to the Sun.

\begin{figure}
\includegraphics[width=13cm]{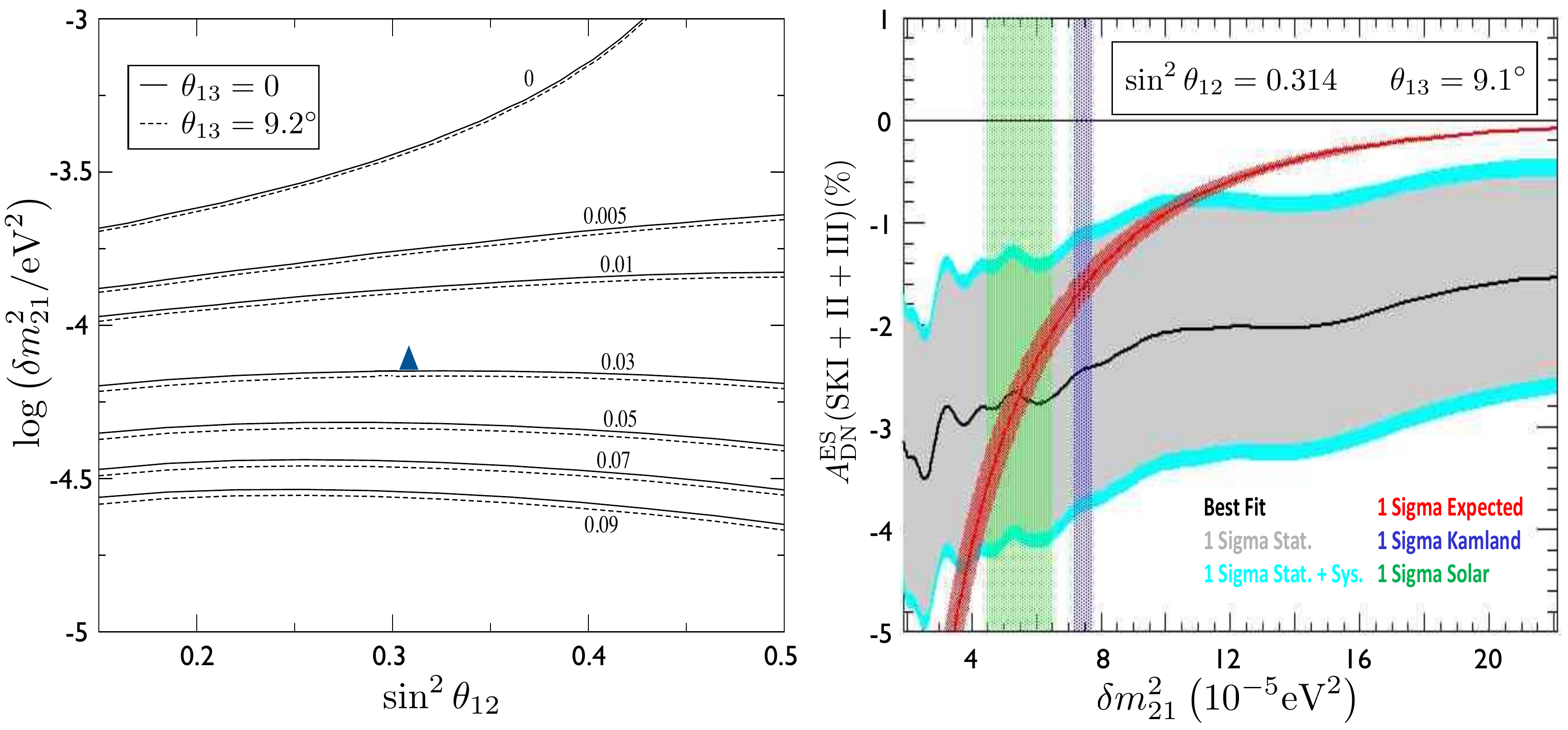}
\caption{(Color online) Left panel: The expected $A_\mathrm{DN}^\mathrm{ES}$ for Super-Kamiokande
as a function of
$\delta m_{21}^2$ and $\sin^2{\theta_{12}}$  for two (solid lines) and three (dashed lines, $\theta_{13}=9.2^\circ$) flavors. The lines are the contours of constant $A_\mathrm{DN}^\mathrm{ES}$.  The blue triangle marks
the best-fit parameters from current global analyses. Adapted from \cite{Blennow04}.
Right panel: $A_\mathrm{DN}^\mathrm{ES}$ values consistent with current values
of $\theta_{12}$ and $\theta_{13}$ (red band), plotted as a function of $\delta m_{21}^2$, compared to SKI+II+III 
results (1$\sigma$ statistical and statistical+systematic errors indicated).  The vertical
bands are the $1\sigma$ KamLAND and solar values for $\delta m_{21}^2$.  Adapted from
A. Renshaw and M. Smy (private communication).}
\label{fig:DNtheory}
\end{figure}

The high counting rate of Super-Kamiokande is an advantage in constraining the contribution to 
$A_\mathrm{DN}$ from the high-energy $^8$B neutrinos.  The expected
 Super-Kamiokande ES $A_\mathrm{DN}$ is illustrated in Fig. \ref{fig:DNtheory}
as a function of $\delta m_{21}^2$ and $\sin^2{\theta_{13}}$, in the two- and three-flavor cases.  For 
the current global best-fit parameters, the effect should be about $-$3\%.  The results for the four phases
\begin{equation}
\begin{array}{llll} -2.1 \pm 2.0 \pm 1.3\% & \mathrm{SK~I} &~~~~~~~6.3 \pm 4.2 \pm 3.7\% & \mathrm{SK~II}\\ -5.9 \pm 3.4 \pm 1.3\% &\mathrm{SK~III} &~~~~-5.2 \pm 2.3 \pm 1.4\%  &\mathrm{SK~IV~(preliminary)} 
\end{array}
\end{equation}
yield a combined result of $A_\mathrm{DN}^\mathrm{ES}(\mathrm{SK}) = -4.0 \pm 1.3 \pm 0.8$, in good agreement with expectations but still consistent
with no effect at 2.6$\sigma$ \citep{Smy}.

The SNO Collaboration has analyzed day-night effects in the $\nu_e$ channel in their combined analysis,
approximating the effect as linear in $E_\nu$,
\begin{equation}
A_\mathrm{DN}^{\nu_e}(\mathrm{SNO}) = - a_0 -a_1 \left( {E_\nu \over 10 \mathrm{~MeV}}-10 \right).
\end{equation}
The null hypothesis that there are no day/night effects influencing the $\nu_e$ survival probability (so $a_0=0,~a_1=0$)
yielded a $\Delta \chi^2 =1.87$ (61\% c.l.) compared to the best fit \citep{SNOCombined}.  The Borexino result for the ES 
at 862 keV \citep{BorexinoDN},
\begin{equation}
A_\mathrm{DN}^\mathrm{ES}(862~\mathrm{keV}) = -(0.001\pm 0.012\pm 0.007),
\end{equation}
is consistent with the expectation that $|A_\mathrm{DN}^\mathrm{ES}(862~\mathrm{keV}|
\lesssim 0.001$ for the LMA solution at this energy.

\section{NEUTRINO CONSTRAINTS ON SOLAR STRUCTURE}
\label{sec:six}

One of the important consequences of the increasingly precise understanding of neutrino flavor physics
is the opportunity to return to one of the early goals of solar neutrino spectroscopy, using the neutrino
as a probe of the physics of the solar interior.  Neutrino fluxes, sensitive to nuclear reaction rates and
core temperature, can be combined with helioseismic observations, sensitive to radiative opacities and
microscopic diffusion, to place stringent constraints on the SSM and to test some of its implicit assumptions.
This program is of broad significance to stellar astrophysics because the SSM is a particular application
of the general theory of main-sequence stellar evolution.  Because we know the Sun's properties far
better than those of any other star, the SSM provides one of our best opportunities to test that theory
against precise data, and thus to identify shortcomings.

A decade ago the SSM was in spectacular agreement with observations apart from solar neutrino
data, a fact that supported suggestions that the solar neutrino problem might have a nonsolar origin.   
However, as the SSM makes a number of simplifying assumptions, it is perhaps inevitable that some
experimental test of our Sun will eventually demonstrate the model's shortcomings.
Over the past decade, the development of 3D 
hydrodynamic models of near-surface solar convection, a more careful selection of spectral lines, and,
in some cases, relaxation of the assumption of local thermodynamic equilibrium in line formation have
led to significant changes in the analysis of data on photospheric absorption lines.   The most
recent revisions reduced the abundances of the volatile CNO elements and Ne by $\sim$ 0.10-0.15 dex
(equivalently, by $\sim 25-40$\%),
relative to older compilations of solar abundances -- though significant debate continues.
The differences between the new AGSS09 and the older GS98 abundances can be summarized 
in the respective total-metal-to-hydrogen ratios of (Z/X)$_\odot$ $\sim$ 0.018 and 0.023, respectively.
As the SSM assumes a homogeneous zero-age composition, adoption of the AGSS09
abundances produces a modern Sun with a lower core metallicity, affecting solar neutrino flux predictions
and substantially degrading the agreement between the SSM sound velocity profile and that
deduced from helioseismology.  

\subsection{The Solar Abundance Problem and its SSM Implications}
Past studies of SSMs with low Z/X interiors similar to that of the AGSS09-SFII SSM have revealed a number of
difficulties with respect to observation: the radius of the convective zone boundary $R_{CZ}$, the interior 
sound speed and density profiles, and the surface He abundance Y$_S$  all move outside the ranges
determined from helioseismic analyses \citep{Bahcall04,Basu04,Montalban04}.  These changes reflect the increase in the efficiency of radiative
transport and decrease in core molecular weight found in low-Z models.  
For example, $R_{CZ}$ moves outward in low-Z models because radiative
transport dominates over a larger fraction of the solar interior.  Similarly, Y$_S$ decreases:
As SSM energy generation is fixed by the measured luminosity $L_\odot$, the cooler core in low-Z models
must be compensated by an increase in the available fuel X, and consequently
a lower core Y and thus surface Y$_S$.  In contrast, SSM predictions using the older,
high-Z GS98 abundances are in
much better agreement with observation.  (See Table \ref{tab:SSM} and Fig. \ref{fig:helio}.)  The inconsistency between the SSM parameterized using the best
current description of the photosphere (AGSS09 abundances) and the SSM parameterized to
optimize agreement with helioseismic data sensitive to interior composition (GS98 abundances) is known as
the solar abundance problem.

The solar abundance problem could have a pedestrian solution:  the 3D analysis of \cite{Caffau10,Caffau09,Caffau08} yielded abundances higher than the AGSS09 values, though this appears to
be due to spectral line choices rather than photospheric model differences \citep{Grevesse11}, a conclusion
supported by a recent comparison between solar model atmospheres computed by different groups
\citep{Beeck12}.  Alternatively,
an upward adjustment in associated atomic opacities could compensate for a low-Z interior, if some 
justification for such a change could be identified.

However, the solar abundance problem could be more fundamental.  An important
assumption of the SSM -- a homogeneous zero-age Sun -- is not based on observation, but instead
on the theoretical argument that the proto-Sun likely passed through a 
fully convective Hayashi phase as the pre-solar gas cloud collapsed, thereby destroying any composition
inhomogeneities that might have existed.   Yet we know that chemical inhomogeneities were
re-established during solar system formation: processes operating in the proto-planetary disk removed 
 $\sim 40-90 M_\oplus$ of metal from the gas, incorporating this material in the gaseous giants \citep{guillot05}.  The
 gas from which these metals were scoured -- representing perhaps the last $\sim$ 5\% of
 that remaining in the disk --
 would have been depleted in metal, and enriched in H and He. The fate of that gas is unknown, but if
 it were accreted onto the Sun, it plausibly could have 
 altered the composition of the convective zone, depending on the timing of the accretion and thus the
 maturity of the proto-Sun's growing, chemically segregated radiative core.
That is, as there is a candidate mechanism for altering the convective zone late in proto-solar evolution,
involving enough metal to account for the AGSS09/GS98 differences, it is not obvious
that the SSM assumption of homogeneity is correct.

The solar abundance problem has three connections to neutrinos:
\begin{enumerate}
\item Neutrino fluxes are sensitive to metallicity, and thus can be used to cross-check the conclusions drawn
from helioseismology.  Below we describe what the neutrino fluxes currently tell us.
\item Neutrino fluxes place important constraints on nonstandard solar models (NSSMs) motivated
by the solar abundance problem, such as those
recently developed to explore accretion from the proto-planetary disk.
\item Planned measurements \citep{SNO+,BorexinoCN} of the CN solar neutrino flux have the potential to directly
measure the solar core abundance of C+N with a precision that will impact the solar abundance problem.
\end{enumerate}

In principle the temperature-dependent $^8$B and $^7$Be neutrino fluxes have sufficient sensitivity to 
metallicity to impact the solar abundance debate.  As Table \ref{tab:fluxes} shows, the AGSS09-SFII and
GS98-SFII SSMs differ by 21.6\% and 9.6\% in their $^8$B and $^7$Be flux predictions, respectively, 
which one can compare to the 14\% and 7\% SSM uncertainties on these fluxes obtained by varying SSM input
parameters according to their assigned errors.   These
total SSM uncertainties were determined by adding in quadrature the individual
uncertainties from 19 SSM input  parameters, including abundance uncertainties
as given in the respective solar abundance compilations.  In the case of the $^8$B flux, the
important uncertainties include those for the atomic opacities (6.9\%), 
the diffusion coefficient (4\%), the nuclear S-factors for $^3$He+$^4$He (5.4\%) and $^7$Be+p (7.5\%),
and the Fe abundance (5.8\%).

Unfortunately the current neutrino data do not favor either abundance set.
The last entries in Table \ref{tab:fluxes}
give the $\chi^2$ functions and compatibility functions $P^\mathrm{agr}$ obtained in the SSM
for the two sets,
following \cite{GGMS10} but using the updated Solar Fusion II  S-factors.  The two models are
identical in quality of fit to the data, with $\chi^2_\mathrm{AGSS09-SFII} = 3.4$ and $\chi^2_\mathrm{GS98-SFII} = 3.5$.
From Table \ref{tab:fluxes} one sees that a SSM model intermediate in metallicity between AGSS09 and GS98
would optimize the agreement.

\subsection{Solar Models with Accretion}
\cite{Guzik06}, \cite{Castro07}, and \cite{Guzik10} considered the possibility that the solar
abundance problem might be due to accretion of metal-poor matter onto the Sun's convective 
envelope, diluting its pre-solar composition. \cite{HS08} suggested the mechanism for such dilution
described above, accretion of disk gas from which the metals were previously
scoured in the process of planetary formation.  Evidence supporting the 
hypothesis that planet formation
could affect the surface metallicity of the host star was offered by \cite{Melendez}, who found that
the peculiar differences in the surface abundances 
of the Sun, measured relative to similar stars (solar twins), correlate with condensation temperatures,
and thus plausibly with disk chemistry.  \cite{Nordlund} argued this accretion scenario might also provide a
natural explanation for the anomalous metallicity of Jupiter.

While the process of planetary formation is not well understood,
the standard picture invokes a chemically differentiated thin disk, with dust, ice, and thus metals concentrated
in the midplane, and outer surfaces dominated by H and He.   This configuration arises from a combination
of gas cloud collapse that is rapid at the poles but inhibited by angular momentum at the equator, and
gas cooling that allows differential condensation of various elements as ice or grains according to their
condensation temperatures.  The formation of planetesimals in the midplane and their self-interactions lead to the
formation of rocky cores of planets.  The gas giants, which are sufficiently
distant from the Sun that ice can augment their cores, reach
masses where tidal accretion of gas can further feed their growth.

The implications of this disk chemistry for solar initial conditions is difficult to assess because of a number
of uncertain parameters.   While we know the planets are substantial metal reservoirs, we do not know
the fate of the depleted gas that dominates the disk surface: it might have been removed by the solar wind, or 
alternatively deposited on the Sun through magnetospheric mechanisms, such as those operating in
young accreting T Tauri stars.  If accretion occurs, its timing relative to early solar evolution is critical.
The timescale for planetary formation is generally estimated at 1-10 M.y., while the SSM 
predicts that the early Sun's convective
boundary moves outward, in response to the growing radiative core, 
over a longer period of $\tau_{CZ}  \sim$ 30 M.y.  If accretion occurs early in this period, when most of
the Sun's mass is in its convective envelope, any resulting nonuniformity in solar abundances would be
negligible.  But if accretion occurs later when the convective zone is similar to that of the modern Sun, and
thus contains only $\sim$ 2\% of the Sun's mass, the chemical processing of $\sim 0.05 M_\odot$ of 
gas, dust, and ice in the planetary nebula could have very significant consequences.   Alternatively,
as suggested by hydrodynamic simulations of the contraction of the protosolar nebula \citep{Wuchterl01}
and stellar models that include episodic accretion during the pre-main-sequence phase with large
mass accretion rates \citep{Baraffe10}, the Sun may have avoided the fully convective phase altogether.
Finally, the mass and
composition of the accreted material is highly uncertain.  Because condensation temperatures for the
elements vary widely -- e.g., from $\sim$1400 K for Fe to $\sim$300 K for C -- the composition would likely 
evolve with time as the disk cools.

Motivated in part by such considerations, a NSSM was recently
developed to test whether early accretion from a planetary disk could resolve the solar abundance 
problem \citep{Serenelli11}.  The work illustrates the importance of
neutrino physics in limiting NSSMs.   In SSMs the pre-solar composition parameters
Y$_\mathrm{ini}$ and Z$_\mathrm{ini}$ as well as the mixing length parameter $\alpha_\mathrm{MLT}$
are determined by iterating the model to reproduce the present-day solar luminosity $L_\odot$, radius $R_\odot$,
and surface metal-to-hydrogen ratio Z$_S$/X$_S$.  This algorithm was modified for the accreting NSSM by allowing
for a mass $M_\mathrm{ac}$ of accreted material with fixed composition (X$_\mathrm{ac}$, Y$_\mathrm{ac}$,
Z$_\mathrm{ac}$), deposited on the early Sun uniformly, beginning at time $\tau_\mathrm{ac,i}$ and lasting
a time $\Delta \tau_\mathrm{ac}$.   Prior to time $\tau_\mathrm{ac,i}$ the Sun was evolved as a SSM with 
mass $M_\odot-M_\mathrm{ac}$ and composition
defined by (X$_\mathrm{ini}$, Y$_\mathrm{ini}$, Z$_\mathrm{ini}$).   During the subsequent
accretion phase, the
simplifying assumption X$_\mathrm{ac}$/Y$_\mathrm{ac}$ $\equiv$ X$_\mathrm{ini}$/Y$_\mathrm{ini}$ was made.
For a given set of fixed accretion parameters, Y$_\mathrm{ini}$, Z$_\mathrm{ini}$,
and $\alpha_\mathrm{MLT}$ were then adjusted iteratively in order to  reproduce $L_\odot$, $R_\odot$,
and Z$_S$/X$_S$, as in the SSM.  Thus Y$_\mathrm{ini}$, Z$_\mathrm{ini}$,
and $\alpha_\mathrm{MLT}$ become functions of the assumed accretion parameters, so that in place of
a single SSM solution, a family of solutions is obtained with different interior compositions. 
Unphysical solutions are rejected, e.g., the AGSS09 Z$_S$/X$_S$ is not compatible
with the accretion of large quantities of metal-rich material at late times, when the convective envelope is thin.

\begin{figure}
\begin{center}
\includegraphics[width=14cm]{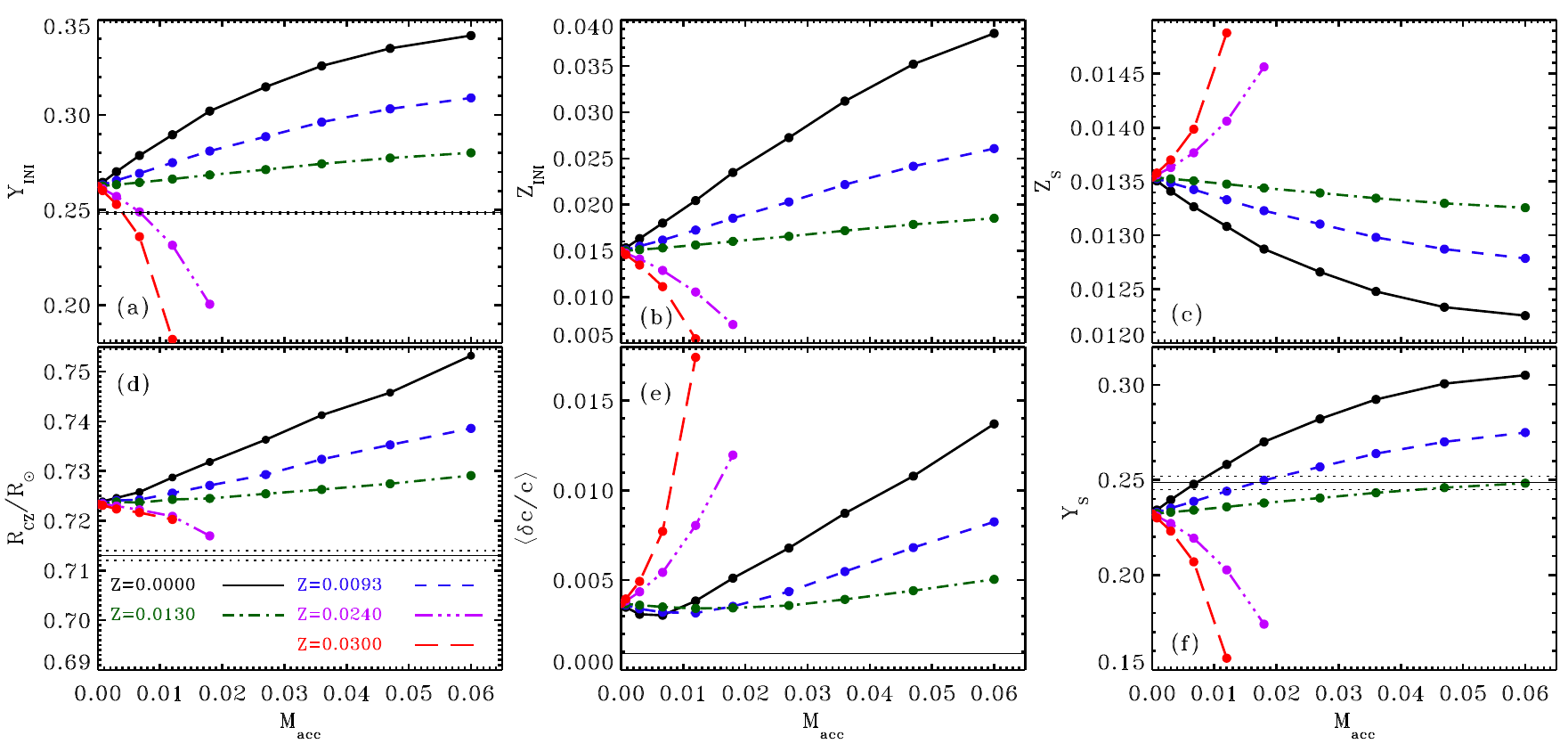}
\caption{(Color online) Accretion on a young Sun with a thin
convective envelope as a function of the
mass and metallicity of the accreted material.  The upper panels show the deduced He and 
metal content of the pre-solar gas, Y$_\mathrm{ini}$ and Z$_\mathrm{ini}$, and present-day Z$_S$,
determined from the luminosity, radius, and AGSS09 photospheric Z$_S$/X$_ S$=0.0178 constraints.  Truncated
trajectories indicate the absence of a physical solution.  The lower panels compare the resulting helioseismic
properties of the models to observation (horizontal bands).} 
\label{fig:accretion_helio}
\end{center}
\end{figure}

Fig. \ref{fig:accretion_helio} shows the helioseismic consequences of late accretion -- accretion onto a young Sun
with a thin convective envelope similar to that of the modern Sun.   Also shown are the
variations in Y$_\mathrm{ini}$ and Z$_\mathrm{ini}$, the initial core He and heavy-element mass fractions,
that can be achieved through accretion.  The helioseismic observables are very constraining.  The candidate
accretion solution to the solar abundance problem that one might naively envision -- a low-Z surface
consistent with AGSS09 and an interior similar to GS98, with higher Z$_\mathrm{ini} \sim 0.019$ and Y$_\mathrm{ini} \sim 0.28$ -- 
can be achieved with metal-free and metal-poor
accretion involving modest accreted masses $M_\mathrm{ac} \sim 0.01 M_\odot$.  These models bring $Y_S$ into good
agreement with observation and produce some improvement in the sound-speed figure-of-merit $\langle \delta c/c \rangle$.
But the lower Z$_S$ that accompanies metal-free/metal-poor accretion allows the convective zone radius to move outward, 
exacerbating the existing AGSS09-SFII SSM helioseismic discrepancy in $R_S$.  

\begin{figure}
\begin{center}
\includegraphics[width=14cm]{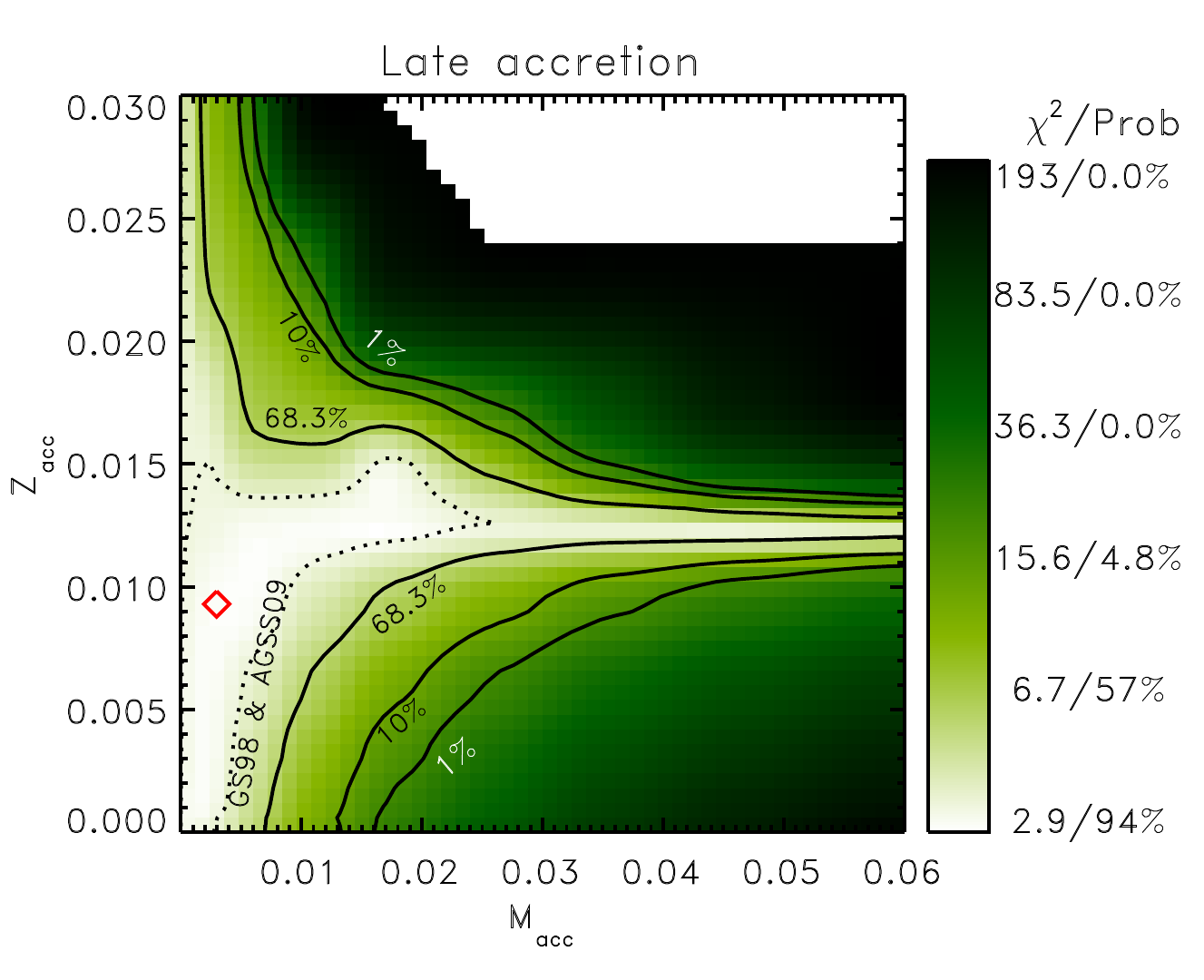}
\caption{(Color online) Global $\chi^2$ solar neutrino analysis for the late accretion NSSM scenario discussed in the text.
Contours are shown for 1\%, 10\% and 68.4\% confidence, with the best-fit model indicated by the red diamond.  The 
fixed-$\chi^2$ contours corresponding to the AGSS09-SFII and GS98-SFII SSM fits are overlaid, showing the very limited 
parameter space where NSSMs with accretion do better than SSMs.} 
\label{fig:accretion_nu}
\end{center}
\end{figure}

Neutrino flux measurements impose a second class of constraints on accreting NSSMs.  The scenario
discussed above -- metal-free or metal-poor accretion with $M_\mathrm{ac} \sim 0.01 M_\odot$ -- can marginally
improve the agreement with experiment, consistent with earlier observations that this produces an
interior similar to GS98, and that the GS98 and AGSS09 SSMs yield equivalent fits to the solar neutrino
data.  But with larger $M_\mathrm{ac}$ the agreement quickly deteriorates as the resulting high-Z interior
leads to rapid increases in the $^8$B and $^7$Be neutrino fluxes.  The power of contemporary neutrino
flux measurements to constrain NSSMs is quite remarkable.
Large classes of accretion parameters -- mass, time, composition, and duration --  lead to modern
Suns with the proper luminosity and radius and the AGS009 Z$_S$, while
satisfying the underlying equations of stellar structure.
Yet very few of these solutions produce acceptable neutrino fluxes, as Fig. \ref{fig:accretion_nu} illustrates.  

Effectively the recent progress made on neutrino mixing angles and mass differences has made the
neutrino into a well understood probe of the Sun.   We now have two
precise tools, helioseismology and neutrinos, that can be used to see into the solar interior,
complementing the more traditional astronomy of solar surface observations.  Effectively we have come
full circle: the Homestake experiment was to have been a measurement of the solar core temperature,
until the solar neutrino problem intervened.

\subsection{Neutrinos as a Direct Probe of Solar Composition: SNO+}
While solar fusion is dominated by the pp chain, the CN cycle generates $\sim$ 1\% of solar energy as well as
$^{13}$N and $^{15}$O neutrino fluxes of 2.96 (2.17) and 2.23 (1.56) $\times 10^8$/cm$^2$s, respectively, for the
GS98-SFII (AGSS09-SFII) SSM.  These fluxes can be measured in scintillation detectors in an energy window of 800-1400 keV,
provided backgrounds in the detectors are reduced to acceptable levels, including in particular interference from
the decay of $^{11}$C produced by penetrating muons.  Borexino has already established a limit on the CN fluxes,
and is pursuing strategies to improve the measurement by vetoing interfering backgrounds \citep{BorexinoCN}.  A new scintillation detector under construction 
in the cavity previously occupied by SNO will have three times the volume of Borexino and 1/70th the muon
background, due to SNOLab's two-kilometer depth \citep{SNO+}.  A SNO+ CN neutrino signal/$^{11}$C background of $\sim$ 10 is expected.

A measurement of the CN neutrinos would test our understanding of hydrogen fusion
as it occurs in main-sequence stars substantially more massive than the Sun. It
could also play an important role in the solar abundance problem.

The CN neutrino fluxes, like the $^8$B and $^7$Be neutrino fluxes, depend sensitively on core temperature $T_C$ and 
thus respond to variations in SSM input parameters that affect $T_C$.
But in addition, the CN fluxes have a linear dependence on the pre-solar core
abundances of C and N that is unrelated to $T_C$, reflecting the proportionality of CN neutrino fluxes to the abundances that catalyze the
hydrogen burning.  The $T_C$ power-law relationships for the CN and $^8$B
neutrino fluxes and the additional linear dependence of the CN neutrino flux on the abundance of C+N can be
exploited to relate solar neutrino measurements to the Sun's pre-solar C and N abundances \citep{HS08}
\begin{eqnarray}
{ \phi(^{15}\mathrm{O}) \over \phi(^{15}\mathrm{O})^\mathrm{SSM}}= \left[{ \phi(^{8}\mathrm{B}) \over \phi(^{8}\mathrm{B})^\mathrm{SSM}}\right]^{0.729}   x_{C+N} ~~~~~~~~~~\nonumber \\
\times  \left[ 1 \pm 0.006 \mathrm{(solar)}\pm 0.027 \mathrm{(D)} \pm 0.099 \mathrm{(nucl)}\pm 0.032\mathrm{(\theta_{12})} \right]
 \label{eq:SNO+}
\end{eqnarray}
where $x_\mathrm{C+N}$ is the C+N number abundance normalized to its SSM value. The uncertainties were derived
from SSM logarithmic derivatives, as described in Sec. \ref{sec:two}.  The first two of these represent
variations in all SSM parameters other than the nuclear cross sections -- including $L_\odot$, the opacity, solar
age, and all abundances other than C and N, using abundance uncertainty intervals of
\begin{equation}
 x_j \equiv 1 \pm \left|{ \mathrm{Abundance}_i^{GS98} - \mathrm{Abundance}_i^\mathrm{AGSS09} \over (\mathrm{Abundance}_i^\mathrm{GS98} + \mathrm{Abundance}_i^\mathrm{AGSS09})/2 } \right| .
 \end{equation}
Apart from the diffusion parameter $D$, the net effect of the variations in these quantities is an uncertainty
of 0.6\%: we have formed a ratio of fluxes that is effectively insensitive to $T_C$.
The diffusion parameter $D$ is an exception because our expression relates 
 contemporary neutrino flux measurements to the pre-solar number densities of C and N, and thus must be
 corrected for the effects of diffusion over 4.6 Gy.  The differential effects of diffusion on the ratio 
 creates an uncertainty of 2.7\%, the only significant nonnuclear solar uncertainty.

Equation (\ref{eq:SNO+}) is written for instantaneous
fluxes, and thus must be corrected for the energy-dependent effects of oscillations.  
The SNO combined analysis
result  $\theta_{12} = 34.06^{+1.16}_{-0.84}$ -- or equivalently the Bari or Valencia global analysis results
of Sec. \ref{sec:five} -- imply a $\lesssim$ 3.2\% uncertainty in the flux comparison of Eq. (\ref{eq:SNO+}).
Finally, there are nuclear physics uncertainties.  These dominate
the overall error budget, with the combined (in quadrature) error reflecting a 7.2\% uncertainty from the $^{14}$N(p,$\gamma$) reaction
and a 5.5\% uncertainty from $^7$Be(p,$\gamma$).  

The SNO and Super-Kamiokande measurements of the $^8$B flux have reached a precision of 3\%.  SNO+ has the potential
to measure the $^{15}$O flux to about 7\% in three years of running.  Assuming such a result from SNO+ and combining all
errors in quadrature, one finds that the pre-solar C+N abundance can be determined to $\pm$ 13\%.  The precision could be improved 
substantially by addressing the uncertainties in the S-factors S$_{114}$ and S$_{17}$.  But even without such improvements, the 
envisioned SNO+ result would
have a major impact, given the existing $\sim$  30\% differences in the AGSS09 and GS98 C and N
abundances.  

\section{OUTLOOK AND CHALLENGES}
\label{sec:seven}
In this review we have summarized the results of 50 years of work on solar neutrinos.  
The field has been characterized by very difficult experiments carried out with great success,
producing results fundamental to two standard models, our standard theory 
of stellar evolution and our standard model of particle physics.  Thirty years of debate over
the origin of the solar neutrino problem -- a misunderstanding of the structure of the Sun, or an incomplete
description of the properties of the neutrino -- ended with the discovery of massive
neutrinos, flavor mixing, and MSW distortions of the solar neutrino spectrum.  The quest 
to resolve the solar neutrino problem spurred the development of a new generation of
active detectors of unprecedented volume and radiopurity -- SNO, Super-Kamiokande,
and Borexino -- that have made solar neutrino spectroscopy a precise science. This
technology has opened up new possibilities.

The program of solar neutrino studies envisioned by Davis and Bahcall has been only partially
completed.  Borexino has extended precision measurements to low-energy solar neutrinos,
determining the flux of $^7$Be neutrinos to 5\%, and thereby confirming the expected increase in the
$\nu_e$ survival probability for neutrino energies in the vacuum-dominated region.  First results
on the pep neutrino flux have been obtained, as well as a limit on the CN neutrino fluxes.  
But a larger, deeper version of Borexino, SNO+, will likely be needed to 
map out the low-energy solar neutrino spectrum in detail, including the CN neutrino contributions.
The observation of these neutrinos in the Sun would provide an important test of the nuclear
reactions we believe dominate energy generation in massive hydrogen-burning stars.

There are challenging tasks remaining.  The flux of solar pp neutrinos, known theoretically
to $\pm$ 1\%, is our most accurately predicted source of $\nu_e$s.   The detection of these
neutrinos \citep{LENS} would check the luminosity condition so widely used in solar neutrino analyses,
the equivalence of solar energy production as measured in emitted photons to that 
deduced from the neutrinos -- a test also relevant to possible new physics,
such as sterile neutrinos \citep{sterile}.  The pp neutrinos could also provide our most precise
low-energy normalization for calibrating the MSW effects that turn on with increasing
energy.  We have discussed the excellent prospects that SNO+, by measuring the $^{15}$O
CN neutrinos (1.73 MeV endpoint), will directly determine the pre-solar C+N content of the solar core to a precision
that will impact the solar abundance problem.  Though a much more challenging task, the measurement of
the lower energy $^{13}$N neutrinos (1.20 MeV endpoint) to high precision could yield
separate determinations of the core C and N abundances: the $\sim$ 35\% higher flux of
$^{13}$N neutrinos reflects the ongoing burning of pre-solar carbon in the Sun's cooler outer core,
as discussed in Sec. \ref{sec:two}.
Finally, there is the next-to-next-generation challenge of measuring the Sun's thermal neutrinos \citep{HW2000},
a solar source that becomes significant below 10 keV, with a peak
flux density of $\sim10^9/\mathrm{cm}^2/\mathrm{s}/\mathrm{MeV}$.
Neutrinos of all flavors are produced by various $Z_0 \rightarrow \nu \bar{\nu}$
processes operating in the Sun.  Their fluxes are a sensitive thermometer for the solar core,
independent of nuclear physics, and depend
on the core abundance of Fe and other heavy elements, due to the contribution of free-bound
transitions where a $Z_0$ is radiated.  The most likely candidate for detection
may be the $\bar{\nu}_e$s, as the correspondence between thermal neutrino energies
and typical inverse atomic scales suggests some form of resonant absorption process.

In retrospect, it is remarkable that we have learned so much fundamental physics from an
object as manifestly complicated as the Sun -- with all of its 3D physics, magnetic fields,
convective and radiative transport, and nontrivial nuclear and atomic physics.  Apart from 
improving the precision of input parameters and small tweaks in the physics, such as the
inclusion of He and heavy element diffusion, the SSM that helped us extract new neutrino physics
from the Sun is the same model Bahcall, Fowler, Iben, and Sears employed in 1963 to
make the first prediction of solar neutrino fluxes.  It would not be surprising to see cracks
finally appearing in that model, under the relentless pressure of experiments measuring 
neutrino fluxes with few percent precision and helioseismic mappings of interior sound
speeds to a few parts in a thousand.  

We have reviewed one possible crack in some detail -- the solar abundance problem. 
Despite the lack of consensus on this issue, it has raised an interesting question:
why have we done so well with a model that assumes a homogeneous zero-age Sun,
when the large-scale segregation of metals is apparent from planetary structure?
The abundance problem has been linked to dynamical issues in the collapse of
the solar nebula, the metallicity of solar twins, and the anomalous composition of Jupiter.
We also have seen that neutrinos could provide a crucial check on the possibility of
an inhomogeneous Sun because of the sensitivity of the CN neutrino flux to the C+N 
content of the core.  It could well turn out that we are encountering the first
evidence that a more complete model is needed -- 
perhaps one that moves beyond an isolated Sun, to address the chemical
and dynamical coupling of the Sun and planets at their births.  Just as the SSM 
allowed us to exploit our best understood star to test the general theory of main-sequence
stellar evolution, an effort to develop a Standard Solar System Model -- one with gas cloud collapse,
disk formation and accretion, realistic pre-solar evolution, the growth of planets, and the 
coupled astrochemistry of the Sun and planets --
could provide a needed template for interpreting what we are now learning about exoplanets
and their host stars. 

We are very grateful to Michael Smy for several helpful discussions on Super-Kamiokande.  This work was supported in part by the US Department of Energy under contracts DE-SC00046548 (UC Berkeley),
DE-AC02-98CH10886 (LBNL),  and 
DE-FG02-97ER41020 (CENPA, Univ. Washington). AS is  partially supported  by 
the  European  Union International
Reintegration Grant  PIRG-GA-2009-247732, the MICINN  grant AYA2011-24704,
the  ESF EUROCORES Program  EuroGENESIS (MICINN  grant EUI2009-04170),  the SGR
grants  of   the  Generalitat  de   Catalunya,  and  EU-FEDER  funds.

\section{LITERATURE CITED}

\bibliographystyle{Astronomy}
\bibliography{solarnu}

\end{document}